\documentclass[prr,twocolumn,aps,showpacs,superscriptaddress,longbibliography]{revtex4-1} 

\usepackage{amsmath}
\usepackage{amssymb}
\usepackage{amsmath}
\usepackage{txfonts}
\usepackage{graphicx} 
\usepackage{epsfig}
\usepackage{float} 
\usepackage[T1]{fontenc}
\usepackage{times}
\usepackage{color} 
\usepackage{cases}
\usepackage{xspace}
\usepackage{amssymb,amsmath}
\usepackage{amsbsy}
\usepackage{braket}
\usepackage{tabularx}
\usepackage{mathtools}          

\usepackage[utf8]{inputenc}

\DeclareMathOperator{\sech}{sech}

\usepackage{ifpdf}
\ifpdf
\usepackage{epstopdf}
\fi

\usepackage[unicode]{hyperref}
\hypersetup{
	pdftitle={spinnorBEC},
	pdfauthor={Yu},
	pdfsubject={Vortexlines},
	colorlinks=true,
	linkcolor=blue,
	citecolor=blue,
	filecolor=black,
	urlcolor=blue
}
\def\urlprefix{}
\def\url#1{}

\usepackage{bm}
\usepackage{color}

\newcommand{\be}{\begin{equation}}

\newcommand{\ee}{\end{equation}}

\newcommand{\bea}{\begin{eqnarray}}

\newcommand{\eea}{\end{eqnarray}}

\newcommand{\nn}{\nonumber }

\newcommand{\Tr}{{\rm Tr}}

\begin{document}

\title{Core structure of static ferrodark solitons in a spin-1 Bose-Einstein condensate}

\author{Xiaoquan Yu}
\email{xqyu@gscaep.ac.cn}
\affiliation{Graduate School of  China Academy of Engineering Physics, Beijing 100193, China}
\affiliation{Department of Physics, Centre for Quantum Science, and Dodd-Walls Centre for Photonic and Quantum Technologies, University of Otago, Dunedin 9016, New Zealand}

\author{P.~B.~Blakie}

\affiliation{Department of Physics, Centre for Quantum Science, and Dodd-Walls Centre for Photonic and Quantum Technologies, University of Otago, Dunedin 9016, New Zealand}

\begin{abstract}
We develop an analytical description of static ferrodark solitons, the $\mathbb{Z}_2$ topological defects in the magnetic order, in the easy-plane phase of ferromagnetic spin-1 Bose-Einstein condensates.  We find that the type-I ferrodark soliton has a single width while the type-II ferrodark soliton exhibits two characteristic length scales. The proposed ansatzes show excellent agreement with numerical results.  Spin-singlet amplitudes, nematic tensor densities and nematic currents of ferrodark solitons are also discussed.  The $\mathbb{Z}_2$ topological defects in the mass superfluid order, dark-dark-dark vector solitons, are obtained exactly in the parameter regime where exact ferrodark solitons exist. The dark-dark-dark vector soliton has higher excitation energy than ferrodark solitons.

\end{abstract}

\maketitle

\section{Introduction}

A spin-1 Bose-Einstein condensate (BEC) is a coherent state constituted by atoms occupying hyperfine spin levels $\ket{F=1,m=+1,0,-1}$ that interact via  spin-mixing collisions, exhibiting  superfluidity and magnetic orders~\cite{Ho98,OM98,Ketterle1998, Stampernatrue2006,StamperRMP,KAWAGUCHI12}. This spinor superfluid supports topological excitations that are absent in scalar and two-component BECs,  adding novel features to phenomena such as the Berezinskii-Kosterlitz-Thouless transitions~\cite{James2011,KobayashiBKT} and out-of-equilibrium processes in superfluids,
including Kibble-Zurek mechanism~\cite{KZSaito2007,KZDamski2007}, phase ordering~\cite{Lamacraft2007,Moore2007,Williamson2016a,Williamson2017,symes2017,Schmied2019,Bourges2017} and thermalization processes~\cite{PrethermalizationBarnett2011,Fujimoto2019}. It also provides, additional to magnetic thin films, a prominent platform for exploring quantum magnetism.

In the presence of a magnetic field along the $z$-axis, the ground state of a ferromagnetic spin-1 BEC can be in the easy-axis or broken-axisymmetry ferromagnetic phases, or the non-magnetized polar phase, each separated by a quantum phase transition~\cite{StamperRMP,KAWAGUCHI12}. The easy-plane phase, being the broken-axisymmetry phase with the magnetization lying in the $xy$-plane, is particularly interesting as it describes $XY$ magnetism and supports various unique topological excitations~\cite{polarcorevortexLewis,polarcorevortexTurner,MDWYuBlair,YuBlairmovingMDW}. A substantial amount of work has been devoted to studying quenches from the polar phase to the easy-plane phase by a change in the magnetic field~\cite{Stampernatrue2006,Saito2007pcvformation,Lamacraft2007,Saito2005DW,KZDamski2007,Williamson2016a,Prufer2018a,Gasenzer2019}. Here polar core vortices are the relevant topological defects in two dimensions (2D) and determine the Kibble-Zurek scaling and the universal dynamic scaling in phase ordering at late times~\cite{Williamson2016a}.  Domain wall structures also appear in the early stage of  quenches due to the linear instability, but these domain walls are transient and continually decay to polar core vortices~\cite{Saito2007pcvformation}.
In one-dimension (1D), domain walls/solitons are relevant defects and contribute the Kibble-Zurek scalings~\cite{KZDamski2007,KZSaito2007} and may play an important role in determining the universal scaling behaviors at later times of the quench ~\cite{Prufer2018a,Gasenzer2019} (i.e., the so-called non-thermal fixed point).
However, little is known on 
the properties of the relevant domain walls/solitons in ferromagnetic spin-1 condensates.
 
An important question is what topological excitations are supported in the easy-plane phase of ferromagnetic spin-1 BECs ?  Here the term ``topological" refers to the winding number (2D) or the topological charge (1D) of the order parameter being nonzero.
In a scalar BEC,  vortices (2D) and dark solitons (1D) are the relevant topological excitations. In ferromagnetic spin-1 BECs,  the magnetization serves as the local order parameter and quantifies the magnetic order associated with the $\textrm{SO}(3)$ rotational symmetry breaking.
In the easy-plane phase, the spin vortices (defects associated with the continuous $\textrm{SO}(2)$ symmetry) are polar core vortices and Mermin-Ho vortices~\cite{StamperRMP,KAWAGUCHI12,Mermin-Ho2002,Kudo2015a}, while kinks in the transverse magnetization (defects associated with the discrete $\mathbb{Z}_2$ symmetry) have been uncovered only recently thanks to the discovery of ferrodark solitons (FDSs) (also referred to as magnetic domain walls)~\cite{MDWYuBlair,YuBlairmovingMDW}. 
The FDSs are magnetic kinks that connect transverse magnetic domains with opposite magnetizations, signified by the magnetization $\mathbf{F}$ vanishing at the core and changing its sign across the core. The superfluid density $n$ has a dip but does vanish at the core.  
The FDSs have two types and exhibit several distinct features, including positive inertial mass, stability against snake deformations, and oscillations in a linear potential due to the transition between the two types at the maximum speed~\cite{MDWYuBlair,YuBlairmovingMDW}.  
The static profiles of FDSs have been studied in detail when $q \rightarrow 0$~\cite{MDWYuBlair}, where $q$ is the quadratic Zeeman energy of the atom in the magnetic field. However, at finite $q$, except for exactly solvable cases~\cite{MDWYuBlair,YuBlairmovingMDW}, the core structure of FDSs remain unexplored.

In this paper we study the core structure of static FDSs for finite quadratic Zeeman energies ($q>0$) in the whole easy-plane phase. Accurate ansatzes are proposed and show good agreement with numerical results. In particular, we obtain  the characteristic length scales analytically via consistent asymptotic analysis,  finding that type-I FDS has a single length scale while type-II FDS requires two length scales to describe the width near-the-core and the width away-from-the-core, respectively.  Spin-singlet amplitudes, nematic tensor densities and nematic currents of FDSs exhibit rich structure and 
provide a complete description of static FDSs. The dark-dark-dark vector solitons, which are the $\mathbb{Z}_2$ topological defects in the mass superfluid order, are also discussed.
      
\section{Spin-1 BECs}
The mean-field Hamiltonian density of a  spin-1 condensate reads 
 \bea 
\label{Hamioltonian}
	{\cal H}=  \frac{\hbar^2 \left|\nabla \psi\right|^2 }{2M} +\frac{g_n}{2} |\psi^{\dag}\psi|^2+\frac{g_s}{2} |\psi^{\dag} \mathbf{S} \psi|^2 +q \psi^{\dag} S^2_z \psi,
\eea
where the three-component wavefunction $\psi=(\psi_{+1},\psi_{0},\psi_{-1})^{T}$ describes the coherent atomic field in the three atomic hyperfine states  $\ket{F=1,m=+1,0,-1}$, $M$ is the atomic mass, $g_n>0$ is the density-dependent interaction strength,  $g_s$ is the spin-dependent interaction strength,   and $S_{j=x,y,z}$ are the spin-1 matrices [$\mathbf{S}=(S_x,S_y,S_z)$]~\cite{StamperRMP,KAWAGUCHI12}. 
The spin-dependent interaction terms describe spin-mixing collisions: $\ket{00} \leftrightarrow\ket{+1}\ket{-1}$, originating from the spin-dependence of the s-wave collisions~\cite{StamperRMP}. 
The magnetic field is along the $z$-axis and $q$ denotes the quadratic Zeeman energy.

At the mean-field level, the  dynamics of the field $\psi$ is governed by the  spin-1 Gross-Pitaevskii equations (GPEs) which are obtained via the canonical equation 
$\partial \psi_{m}/\partial t=\delta {\cal H}/\delta (i \hbar \psi^{*}_{m})$:
\bea
\label{spin-1GPE}
i\hbar \frac{\partial \psi_{\pm 1}}{\partial t}	&&=\left[H_0+g_s\left(n_0+n_{\pm 1}-n_{\mp 1}\right)+q \right]\psi_{\pm 1}+g_s \psi^2_0 \psi^{*}_{\mp 1} ,  \\
i\hbar \frac{\partial \psi_0}{\partial t}	&&= \left[H_0 +g_s\left(n_{+1}+n_{-1}\right) \right]\psi_0 + 2g_s \psi^{*}_0\psi_{+1}\psi_{-1}, 
\label{spin-1bGPE}
\eea  
where $H_0=-\hbar^2\nabla^2/2M +g_n n$, $n=\sum n_m$ is the total number density  and $n_m=|\psi_m|^2$ is the component density.

A distinguishing feature of spin-1 BECs is that the system supports magnetic orders, with  the magnetization $\mathbf{F}\equiv\psi^{\dag} \mathbf{S} \psi=(F_x, F_y,F_z)$ serving as the local order parameter. Here
\bea
F_{x}&=& \psi^{\dag} S_{x} \psi= \frac{1}{\sqrt{2}}\left[\psi^{*}_0(\psi_{+1}+\psi_{-1})+{\rm H.c.}\right],   \\
F_{y}&=& \psi^{\dag} S_{y} \psi=\frac{i}{\sqrt{2}} \left[\psi^{*}_0(\psi_{+1}-\psi_{-1})-{\rm H.c.}\right],\\ 
F_z&=&\psi^{\dag} S_{z} \psi=|\psi_{+1}|^2-|\psi_{-1}|^2. 
\eea
It is convenient to introduce the transverse magnetization as the complex density
\bea
F_{\perp} \equiv F_x+i F_y=\sqrt{2} \left(\psi_0 \psi^{*}_{+1}+\psi^{*}_0\psi_{-1}\right).
\eea
For ferromagnetic coupling $g_s<0$ ($^{87}$Rb,$^7$Li), $|\mathbf{F}|>0$ and for anti-ferromagnetic coupling $g_s>0$ ($^{23}$Na), $\mathbf{F}=0$.  At $q=0$, the system processes $\textrm{SO}(3)$ rotational symmetry and the total magnetization $\mathbf{M}=\int d \mathbf{r} \, {\mathbf F}$ is conserved. In the presence of a magnetic field ($q\neq 0$), the remaining symmetry is $\textrm{SO}(2)$ rotational symmetry, and yields to the conservation of the total magnetization along the $z$-axis $M_z=\int d \mathbf{r} F_z$.  In contrast to binary BECs, a spin-1 BEC is not an incoherent mixture of condensates, i.e., the system does not process $\textrm{U}(1)\times\textrm{U}(1)\times \textrm{U}(1)$ symmetry~\cite{StamperRMP}.

\section {Easy-plane phase} 
For a uniform
ferromagnetic system ($g_s<0$) with total density $n_b$, the energy
density is 
\bea
{\cal H}= \frac{1}{2}g_n n^2_b+\frac{g_s}{2}|\mathbf{F}|^2+q(n_b-n_0).
\label{energy}
\eea
The uniform ground state is found by minimizing the free energy density  ${\cal F} \equiv {\cal H}-p \,F_z$~\cite{Ketterle1998,Zhang2003}, where $p$ is the Lagrange multiplier. In the following we only consider $F_z=0$. Hence for $q>0$ the magnetization prefers to  lie in the transverse plane, realizing an easy-plane ferromagnetic phase~\cite{StamperRMP,KAWAGUCHI12}. 
Plugging the wavefunction
\bea
\psi=\left(\sqrt{n_{+1}}e^{i\theta_{+1}},	\sqrt{n_0}e^{i\theta_0},\sqrt{n_{-1}}e^{i\theta_{-1}}\right)^{T}
\eea
into the energy function Eq.~\eqref{energy}, we obtain  
\bea
E=2g_s n_0(n_b-n_0)\cos^2\left[\theta_0-\frac{\theta_{+1}+\theta_{-1}}{2}\right]+\frac{g_n}{2} n_b^2+q(n_b-n_0). \nn\\
\eea
The energy $E$ reaches the minimum when
\bea
2\theta_0&=&\theta_{+1}+\theta_{-1}, \\
 n_0&=&n^{g}_0=\frac{n_b (1+\tilde{q})}{2},  \\
 n_{\pm 1}&=&n^{g}_{\pm 1}=\frac{n_b (1-\tilde{q})}{4}, 
\eea
where $\tilde{q}\equiv -q/(2g_sn_b)$. 
The ground-state magnetization reads
\bea
F_{\perp}^g(\tau)=n_b\sqrt{1-\tilde{q}^2} e^{ i \tau},
\eea
where $\tau=\theta_0-\theta_{+1}=\theta_{-1}-\theta_0$ describes the orientation of the transverse magnetization [the $\textrm{SO}(2)$ rotational angle about the z axis ($e^{-i \tau S_z}$)], quantifying the magnetic order.
The ground wavefunction can be written as  
\bea
\psi^{g}(\theta,\tau)
=\sqrt{n_b}e^{i \theta}e^{-i \tau S_{z}}\left(\sqrt{\frac{n^{g}_{+1}}{n_b}},\sqrt{\frac{n^{g}_0}{n_b}},\sqrt{\frac{n^{g}_{-1}}{n_b}} \right)^{T},
\eea
where $\theta$ describes the global $\textrm{U(1)}$ phase and quantifies the mass superfluid order. Note that the choice of the global phase $\theta$ is not unique and depends on the realization of the spin rotation in spin states. For instance, for the state $ 
e^{i\theta_0}\left(\sqrt{n^{g}_{+1}}e^{-i\tau},\sqrt{n^{g}_0},\sqrt{n^{g}_{-1}}e^{i\tau}\right)^{T}$, $\theta_0$ can be regarded as the global phase. Rearranging the phases in spin states, it becomes  $e^{i\theta_{+1}}\left(\sqrt{n^{g}_{+1}},\sqrt{n^{g}_0}e^{i\tau},\sqrt{n^{g}_{-1}}e^{2\tau i} \right)^{T}$ and then $\theta=\theta_{+1}$.

The positiveness of component densities $n_{\pm 1}$ requires $q<-2g_s n_b$ which sets the parameter range of the easy-plane phase. Note that for $q<0$, under the constraint $M_z=0$, a fragmented state of mixed $F_z=n_b$ and $F_z=-n_b$ domains has lower energy than any uniform state~\cite{KAWAGUCHI12,StamperRMP}.

%

\section{$\mathbb{Z}_2$ topological defects} 
\label{FDS}
In this section, we discuss $\mathbb{Z}_2$ topological defects in the easy-plane phase of a ferromagnetic BEC. Hereafter we consider 1D systems. 
It is useful to introduce the topological charge to characterize $\mathbb{Z}_2$ defects in the magnetic order 
\bea
Q_{F_{\perp}}=\frac{1}{2|F^{g}_{\perp}|}\int^{\infty}_{-\infty} dx \, \partial_x F_{\perp}(x) 
\eea
and the topological charge to characterize $\mathbb{Z}_2$ defects in the mass superfluid order
\bea
Q_{\psi}=\prod_{m=-1,0,1} Q_{\psi_m}
\eea
where
\bea
Q_{\psi_m}=\frac{1}{2 |\psi^{g}_{m}|}\int^{\infty}_{-\infty} dx\, \partial_x \psi_m(x). 
\eea
The topological charges can be also expressed in terms of relevant phases:   
\bea
Q_{F_{\perp}}=\frac{1}{\pi}\int^{\infty}_{-\infty} dx \, \partial_x \tau(x),
\eea
and
\bea
Q_{\psi}=\frac{1}{\pi}\int^{\infty}_{-\infty} dx\, \partial_x \theta (x). 
\eea
It should be understood that here the results are modulo 2. 
The $\textrm{SO}(2)$ spin rotation $e^{-i \pi S_z}$ and the $\textrm{U}(1)$ gauge transformation $e^{-i \pi}$ play the role of charge conjugation operators and change signs of $Q_{F_{\perp}}$ and $Q_{\psi}$, respectively. Note that the spin rotation $e^{-i \pi S_z}$ does not change sign of $Q_{\psi}$ and the gauge transformation $e^{-i \pi}$ keeps sign of $Q_{F_{\perp}}$ unchanged.

\subsection{$\mathbb{Z}_2$ defects in the magnetic order: ferrodark solitons} 
Let us consider two oppositely magnetized domains with $\tau=0$ and $\tau=\pi$ respectively and search for a magnetic kink $F_{\perp}(x)$ that 
interpolates the two domains, namely
\bea
\hspace{-0.2cm}F_{\perp}(-\infty)=F^{g}_{\perp}(\pi) \,\, \text{and} \,\, F_{\perp}(+\infty)=F^{g}_{\perp}(0)=-F_{\perp}(-\infty).
\label{Fcondition}
\eea
It was found that an Ising-type magnetic kink satisfies the condition~Eq.~\eqref{Fcondition}, signified by $F_{\perp}$ vanishing at the core and changing its sign across the core. This kink was referred to as FDS~\cite{YuBlairmovingMDW}.
Exact FDS solutions of Eqs.~\eqref{spin-1GPE} and ~\eqref{spin-1bGPE} were found at $g_s=-g_n/2$~\cite{MDWYuBlair,YuBlairmovingMDW}:
\bea 
F^{\rm I,II}_{\perp}(x)&=&|F_{\perp}^g| \tanh\left(\frac{x}{2\ell^{\rm I,II}_{\rm ex}}\right), \\
n^{\rm I,II}(x)&=&n_b\left[1-\frac{(1\mp \tilde{q})}{2}\sech^2\left(\frac{x}{2\ell^{\rm I,II}_{\rm ex}}\right)\right],\label{eqnIII}
\eea
where $\ell^{\rm I,II}_{\rm ex}=\hbar/\!\sqrt{2g_n n_b M (1\mp\tilde{q})}$, and the minus and plus signs in front of $q$ are referred to as type-I and type-II FDSs, respectively. Note that $\ell^{\rm II}_{\rm ex}\le \ell^{\rm I}_{\rm ex}$ with the equality holding at $q=0$. The corresponding wavefunctions read 
\bea
\psi^{\rm I}_{\pm 1}(x)=\sqrt{n^g_{\pm 1}} \tanh \left(\frac{x}{2 \ell^{\rm I}_{\rm ex}}\right), \quad
\psi^{\rm I}_0(x)=\sqrt{n^g_0} ;
\label{typeIexact}
\eea
and
\bea
\psi^{\rm II}_{\pm 1}(x)=\sqrt{n^g_{\pm 1}}, \quad  \psi^{\rm II}_0(x)=\sqrt{n^g_0} \tanh\left(\frac{x}{2\ell^{\rm II}_{\rm ex}}\right).
\label{typeIIexact}
\eea 
At the core ($x=0$), the magnetization vanishes [$F^{\rm I,II}_{\perp}(0)=0$] while the superfluid density $n$ is finite [$n^{\rm I,II}(0)\neq 0$].  FDSs are topological defects in the magnetic order but not in the superfluid order, namely $Q_{F_{\perp}}=\pm 1$ [$Q_{F_{\perp}}=-1$ for anti-FDSs $e^{-i \pi S_z} \psi^{\rm I,II}(x)$] while $Q_{\psi}=0$. 

The wavefunctions of FDSs Eqs.~\eqref{typeIexact} and ~\eqref{typeIIexact}  can be written as follows:
\bea
\psi^{\rm I}(x)=\left[|\psi^{\rm I}_{+1}|e^{-i\tau(x)}, \quad|\psi^{\rm I}_0| ,\quad |\psi^{\rm I}_{-1}|e^{i\tau(x)}\right]^{T},\\
\psi^{\rm II}(x)=\left[|\psi^{\rm II}_{+1}|, \quad |\psi^{\rm II}_0| e^{i\tau(x)} ,\quad |\psi^{\rm II}_{-1}|e^{i 2\tau(x)}\right]^{T},
\eea
where $\tau(x)$ is a step function: $\tau(x<0)=-\pi$ and $\tau(x\ge0)=0$.  
Hence type-I and type-II FDSs might be viewed as 1D analogies of polar core vortices and  Mermin-Ho vortices, respectively.

We consider stationary excitations which satisfy $\psi_{+1}=\psi_{-1}$ [upon a $\textrm{SO}(2)$ spin rotation] and the stationary GPEs become
\bea
\label{SGPE1}
0&&=\left[-\frac{\hbar^2}{2M}\frac{d^2}{dx^2}-\mu + 2g_nn_{\pm 1} +(g_n+2g_s) n_0+q\right]\psi_{\pm 1},\\
0&&=\left[-\frac{\hbar^2}{2M}\frac{d^2}{dx^2}-\mu+g_nn_0 +2(g_n+2g_s)n_{\pm 1} \right]\psi_{0\phantom{+}}, 
\label{SGPE2}
\eea
where the chemical potential $\mu=(g_n+g_s)n_b+q/2$ and the wavefunctions $\psi_m$ are chosen to be real. Clearly $\textrm{U(1)}$ gauge transformations and $\textrm{SO}(2)$ spin rotations keep Eqs.~\eqref{SGPE1} and \eqref{SGPE2} unchanged.  At $g_s=-g_n/2$, Eqs.~\eqref{SGPE1} and \eqref{SGPE2} become decoupled 
and admit exact  type-I [Eq.~\eqref{typeIexact}] and type-II [Eq.~\eqref{typeIIexact}] solutions. It is worth mentioning that the exactly solvable point is within the scope of a $^7$Li spin-1 BEC which has been prepared in a regime with a  strong spin-dependent interaction~\cite{Huh2020a}.

\subsection{$\mathbb{Z}_2$ defects in the mass superfluid order: dark-dark-dark vector solitons}

\begin{figure}[htp] 
	\centering
	\includegraphics[width=0.43\textwidth]{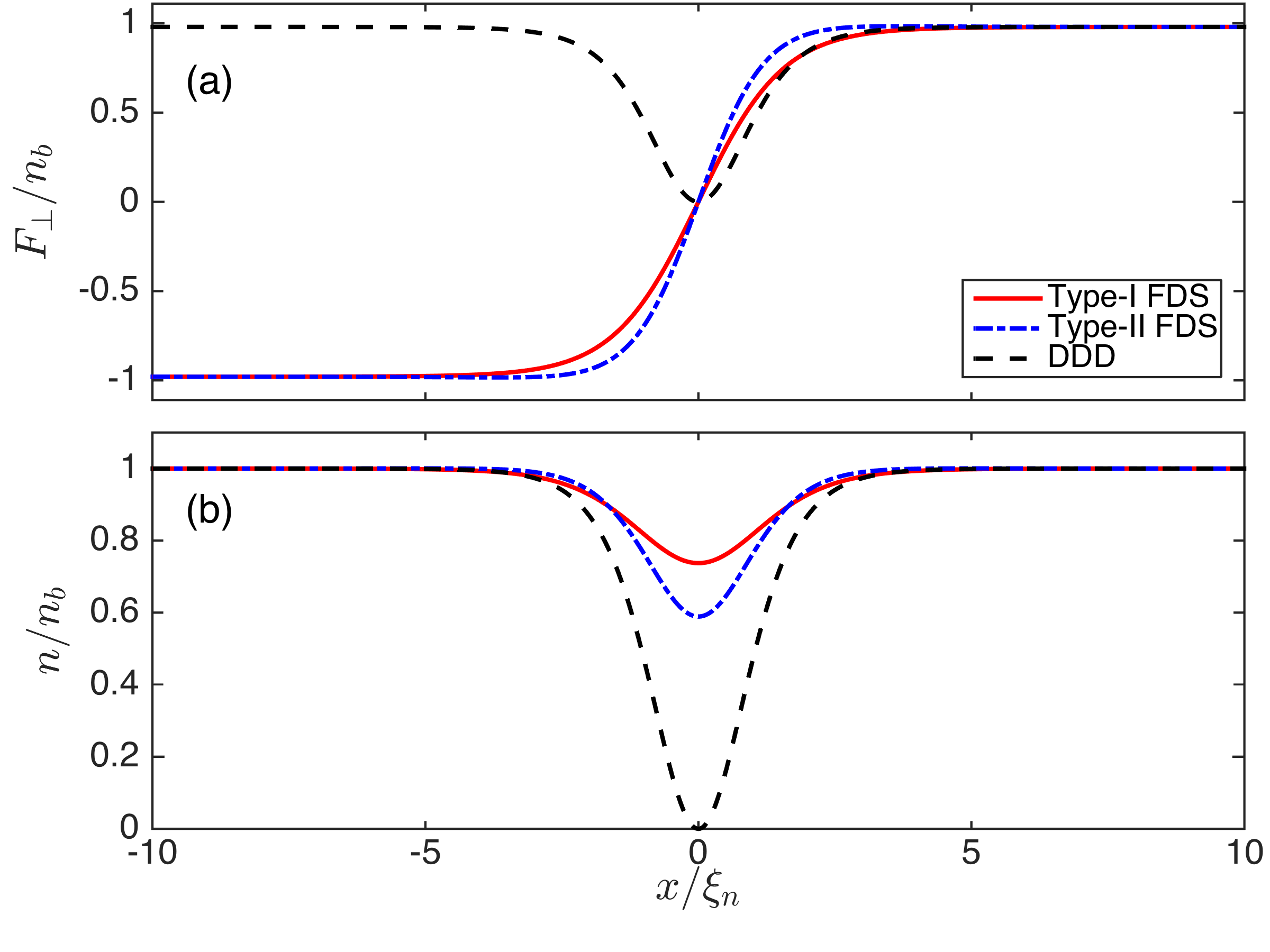}
	\caption{The transverse magnetization  (a) and the density (b) of FDSs and the DDD at $g_s=-0.3g_n$ and $\tilde {q}=0.2$. Here $\xi_n=\hbar/\sqrt{n_b g_n M}$.  \label{f:DDDFDS}} 
\end{figure}

There is another soliton solution $\psi(x)$ to the Eqs.~\eqref{SGPE1} and \eqref{SGPE2} that satisfies
\bea
\psi(-\infty)=\psi^{g}(\pi,\tau) \,\, \text{and} \,\, \psi(+\infty)=\psi^{g}(0,\tau)=-\psi(-\infty),
\label{Scondition}
\eea
and at $g_s=-g_n/2$:	
\bea
\psi^{d}_{\pm 1}(x)=\sqrt{n^g_{\pm 1}} \tanh \left(\frac{x}{2 \ell^{\rm I}_{\rm ex}}\right), \quad
 \psi^{d}_0(x)=\sqrt{n^g_0} \tanh\left(\frac{x}{2\ell^{\rm II}_{\rm ex}}\right). \nn\\
\label{typedexact}
\eea
Here we have chosen that $\tau=0$.
The corresponding transverse magnetization and the total density read 
\bea 
F^{d}_{\perp}(x)&=& |F_{\perp}^g|\tanh\left(\frac{x}{2\ell^{\rm II}_{\rm ex}}\right) \tanh \left(\frac{x}{2 \ell^{\rm I}_{\rm ex}}\right), \\
n^{d}(x)&=&n^b\left[1- \frac{1-\tilde{q}}{2}\sech \left(\frac{x}{2 \ell^{\rm I}_{\rm ex}}\right)^2-  \frac{1+\tilde{q}}{2}\sech\left(\frac{x}{2\ell^{\rm II}_{\rm ex}}\right)^2\right]. \nn\\\label{DDD}
\eea
In the literature this soliton has been referred to as the dark-dark-dark (DDD) vector soliton~\cite{Katsimiga_2021}. 
The DDD vector soliton is the $\mathbb{Z}_2$ topological defect in the mass superfluid order, characterized by changing sign of $\psi^{d}(x)$ across the core [accompanied by $n^{d}(0)=0$] and $Q_{\psi}=1$ [$Q_{\psi}=-1$ for the anti-DDD $e^{-i \pi} \psi^{d}(x)$]. The DDD vector soliton is not a topological defect in the magnetic order as the transverse magnetization $F^{d}_{\perp}$ does not change sign across the core [although $F^{d}_{\perp}(0)=0$] and $Q_{F_{\perp}}=0$. Note that the spin rotation $e^{-i \pi S_z}$ does not change the sign of $Q_{\psi}$, i.e., $Q_{\psi}=1$ for $e^{-i \pi S_z}\psi^{d}(x)$ and $Q_{\psi}=-1$ for $e^{-i \pi S_z}e^{-i \pi}\psi^{d}(x)$. Typical profiles  of  FDSs and the DDD vector soliton are shown in Fig.~\ref{f:DDDFDS}.

\begin{figure}[htp] 
	\centering
	\includegraphics[width=0.46\textwidth]{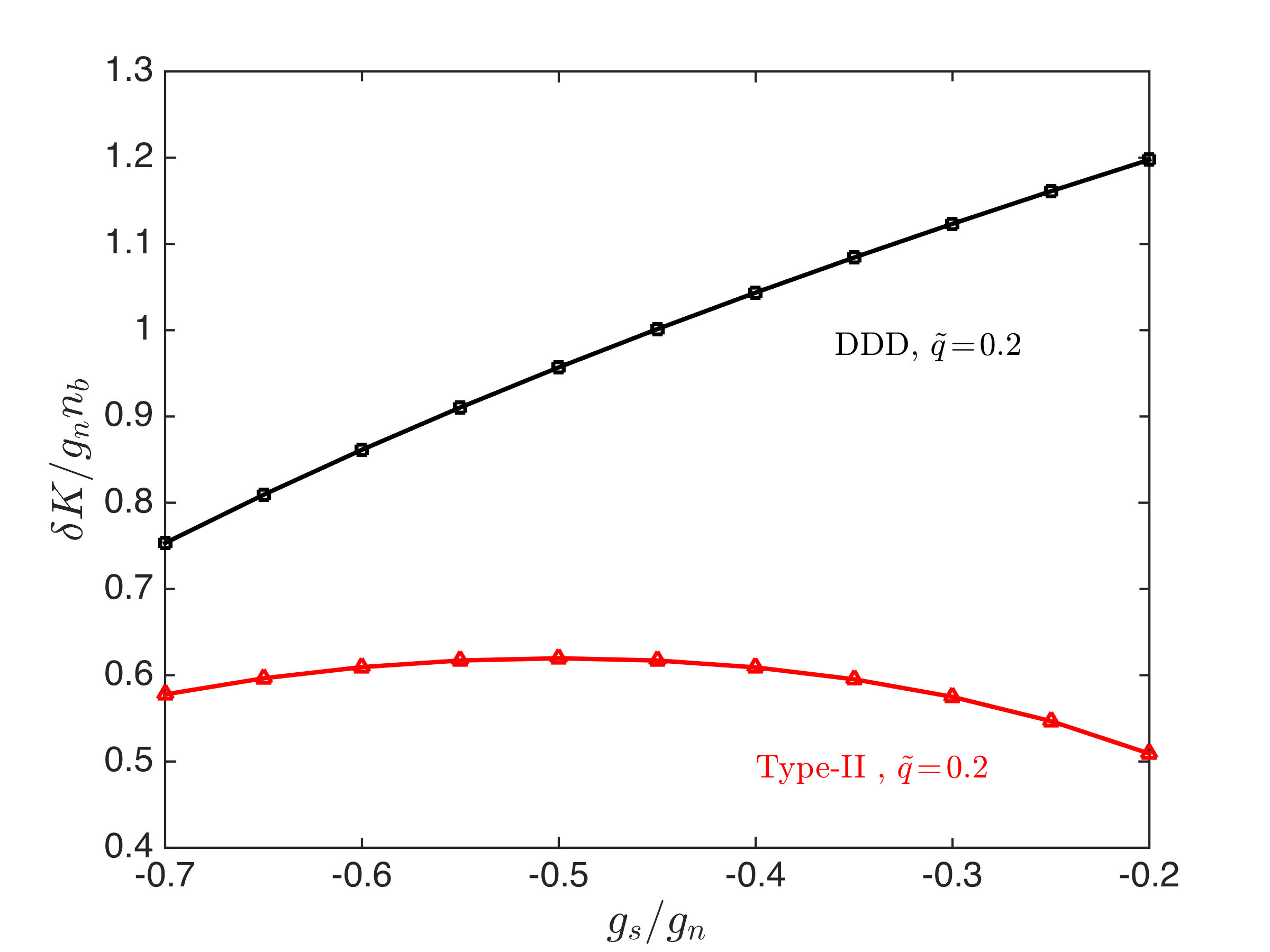}
	\caption{Numerical results for the energies of type-II FDSs (red triangles) and DDD solitons (black squares) at $\tilde {q}=0.2$. Here only the type-II FDS energy curve is presented as the type-I FDS always has lower energy (see Sec.~\ref{excitationenergy}).   \label{f:energy3type}} 
\end{figure}

The energy of a DDD vector soliton is higher than a FDS in the parameter region where the DDD vector soliton is supported~\cite{footnoteDDD} (see Fig.~\ref{f:energy3type}). Here the numerical solutions are obtained using a gradient flow method~\cite{Lim2008a,Bao2008a}.  We quantify the soliton energy as the excess grand canonical energy over that of the ground state, i.e.~as $\delta K \equiv K[\psi]-K_g$, where
\bea
K[\psi]=\int dx \, \left({\cal H}[\psi]-\mu n_b\right),\label{solitonenergy}
\eea
and 
\bea
K_g=\int dx \, \left({\cal H}[\psi^g]-\mu n_b\right)= \frac{n^2_b[g_s (\tilde{q}^2-1)-g_n]}{2}L.
\eea
Here $L$ is the system size. 
 
In the following, we focus on analyzing the core structure of FDSs away from the exactly solvable regime.

\section{Core structure for a finite quadratic Zeeman shift}

The fundamental properties of FDSs do not qualitatively change away from the exactly solvable regime ($g_s/g_n=-1/2$, $0<q<-2g_s n_b$). However, the characteristic length scales do vary as  $g_s$ changes and have been characterized in the $q\rightarrow 0$ limit in Ref.~\cite{MDWYuBlair}.   
This limit allows us to connect, through $\textrm{SO}(3)$ spin rotations, various different degenerate forms of the FDS. At $q=0$, the core structure was conveniently analyzed using a particular rotated state referred to as the sine-Gordon representation.
For $q>0$, the $\textrm{SO}(3)$ symmetry is broken by the magnetic field and the sine-Gordon like solitons are no longer stationary solutions~\cite{MDWYuBlair}.  A different scheme is therefore required to obtain the widths of FDSs.

\subsection{Type-I FDSs} 
\begin{figure}[htp] 
	\centering
	\includegraphics[width=0.43\textwidth]{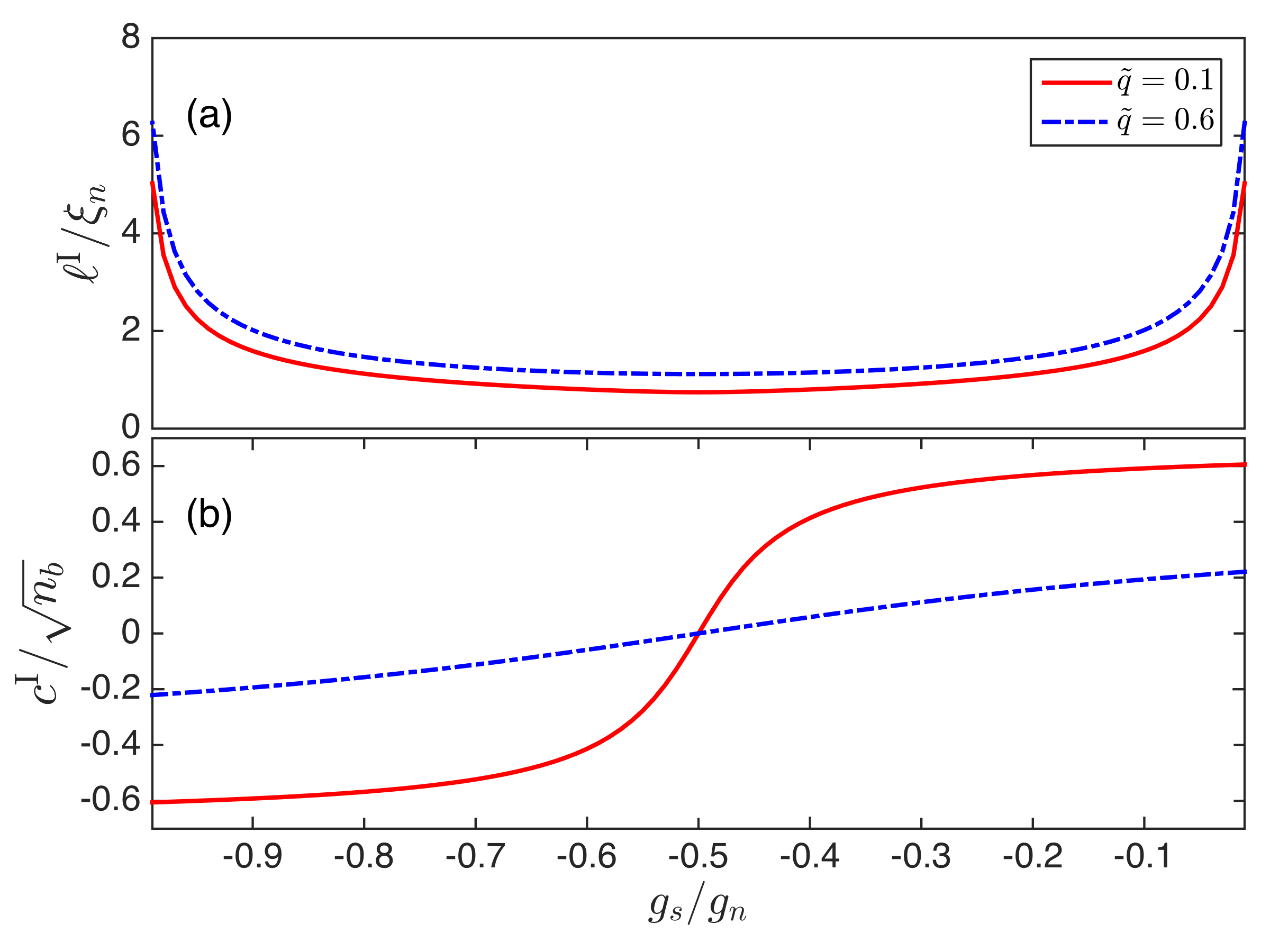}
	\caption{Key parameters in the ansatz of type-I FDSs given in Eqs.~\eqref{typeI1} and \eqref{typeI} as functions of $g_s$ for $\tilde{q}=0.1$ and $\tilde{q}=0.6$.  \label{f:parameterI} } 
\end{figure}

\begin{figure}[htp] 
	\centering
	\includegraphics[width=0.479\textwidth]{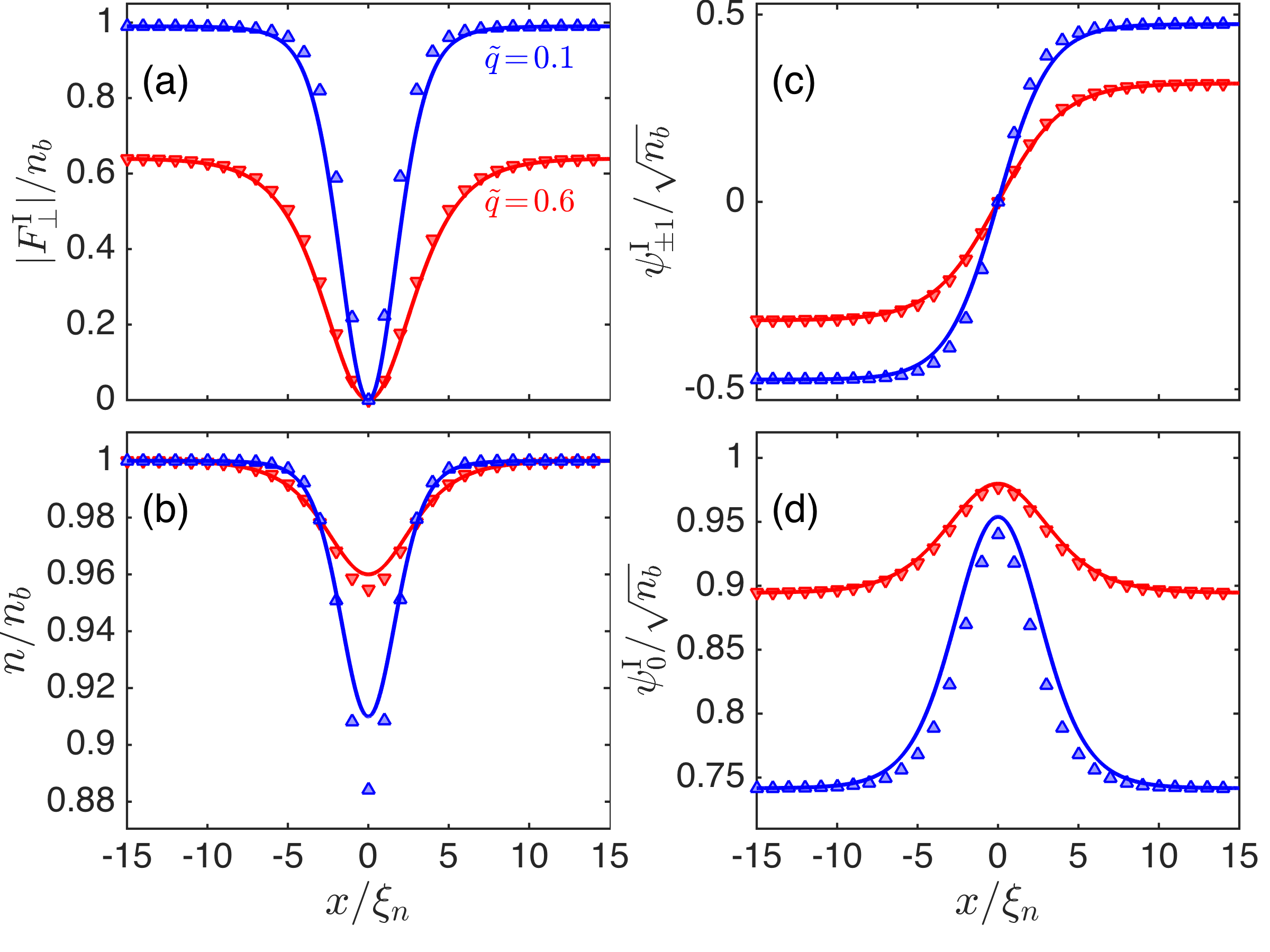}
	\includegraphics[width=0.479\textwidth]{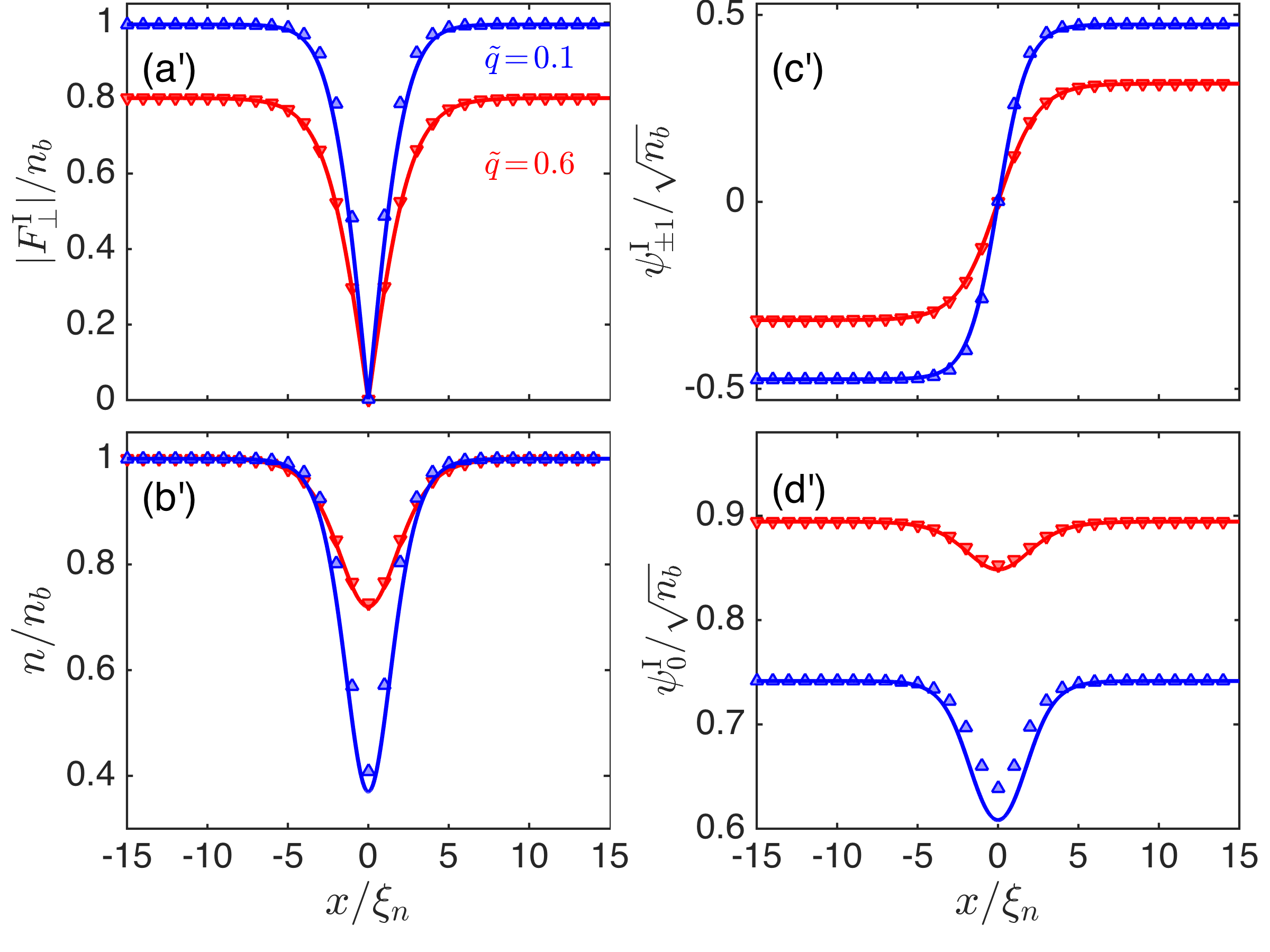}
	\caption{A comparison between analytical predictions of type-I FDSs (solid lines) and numerical results (markers) 
		at $\tilde{q}=0.1$ and $\tilde{q}=0.6$. 	(a), (b), (c) and (d) are for $g_s/g_n=-0.1$ ; (a$'$), (b$'$), (c$'$) and (d$'$) are for $g_s/g_n=-0.7$. \label{f:profileI}} 
\end{figure}

Away from the exactly solvable regime, we propose the following ansatz for type-I FDSs
\bea
\psi^{\rm I}_{\pm 1}(x)&&=\sqrt{n^g_{\pm 1}} \tanh \left(\frac{x}{2 \ell^{\rm I}}\right), \label{typeI1}\\
\psi^{\rm I}_0(x)&&=\frac{c^{\rm I}}{g^{\rm I}+\cosh \left(\frac{x}{\ell^{\rm I}}\right)}+\sqrt{n^g_0},
\label{typeI}
\eea
where $\ell^{\rm I}$ is a length scale describing the core size and $g^{\rm I}$ is introduced to adjust the core structure. 
For $x \gg \ell^{\rm I}$, $\psi_{\pm 1}(x)\sim \sqrt{n^g_{\pm 1}}(1-2 e^{-x/\ell^{\rm I}})$ and $\psi^{\rm I}_0(x) \sim 2 c^{\rm I} e^{-x/\ell^{\rm I}}+\sqrt{n^g_0}$. This ansatz solves  Eqs.~\eqref{SGPE1} and~\eqref{SGPE2} asymptotically ($x \gg \ell^{\rm I}$) if $\ell^{\rm I}$  is a physically acceptable (real and positive) root of the polynomial equation 
\bea
4 M^2 (g_n+g_s) (4 g^2_s n_b-q^2) y^4 +4 g_n g_s  M n_b \hbar ^2 y^2-g_s \hbar ^4=0 \nn \\
\label{widthtypeI}
\eea
and
\bea
c^{\rm I}=-\frac{ (g_n+2 g_s) (2 g_s n_b+q) \sqrt{2 n_b-\frac{q}{g_s}}}{4 \sqrt{g_s  \left[g_s n_b^2 (g_n+2 g_s)^2-q^2 (g_n+g_s)\right]}+2 g_n q }.
\eea
Two positive roots of Eq.~\eqref{widthtypeI} are 
\bea
y_{\pm}=\sqrt{\frac{g_s \hbar ^2}{2 M \left(g_n g_s n_b\pm\sqrt{g_s[g_s n_b^2 (g_n+2 g_s)^2-q^2 (g_n+g_s)]}\right)}}. \nn\\
\eea
It is easy to check that $y_{+}|_{g_s=-g_n/2}=\ell^{\rm I}_{\rm ex}$ [where $\ell^{\rm I}_{\rm ex}$ was introduced after Eq.~(\ref{eqnIII})], hence we identify 
\bea
\ell^{\rm I}=y_{+}.
\eea 
As $g_s\rightarrow -g_n/2$, 
$c^{\rm I}\rightarrow 0$. Moreover $c^{\rm I}$ changes the sign when $g_s$ crosses the exactly solvable point $-g_n/2$ [Fig.~\ref{f:parameterI}(b)], inducing a hump or dip in $\psi_0$ [Fig.~\ref{f:profileI} (d) and (d$'$)].

Near the core $x\sim 0$, solving Eqs.~\eqref{SGPE1} and ~\eqref{SGPE2} to leading order, we obtain 
\bea
g_n \left(\frac{c^{\rm I}}{g^{\rm I}+1}+d\right)^3-\mu  \left(\frac{c^{\rm I }}{g^{\rm I}+1}+d\right)+\frac{2c^{\rm I} \hbar ^2}{4 (g^{\rm I}+1)^2 (\ell^{\rm I})^2 M}=0, \nn\\
\label{coreI}
\eea
which determines the value of $g^{\rm I}$.  Neglecting the last term of Eq.~\eqref{coreI}, we obtain
\bea
g^{\rm I}=\frac{c^{\rm I}}{\sqrt{\mu/g_n}-\sqrt{n^{g}_0}}-1.
\label{gfactor}
\eea
Evaluating the last term in Eq.~\eqref{coreI} using  the result in Eq.~\eqref{gfactor} verifies that it is small, justifying the approximation used here.

The results in Fig.~\ref{f:profileI} compare the ansatz in Eqs.~\eqref{typeI1} and \eqref{typeI} against numerical solution of the GPEs. This comparison shows that the ansatz we have developed provides a very good description of the core structure of the soliton over a wide range of parameters.


\subsection{Type-II FDSs}
\begin{figure}[htp] 
	\centering
	\includegraphics[width=0.43\textwidth]{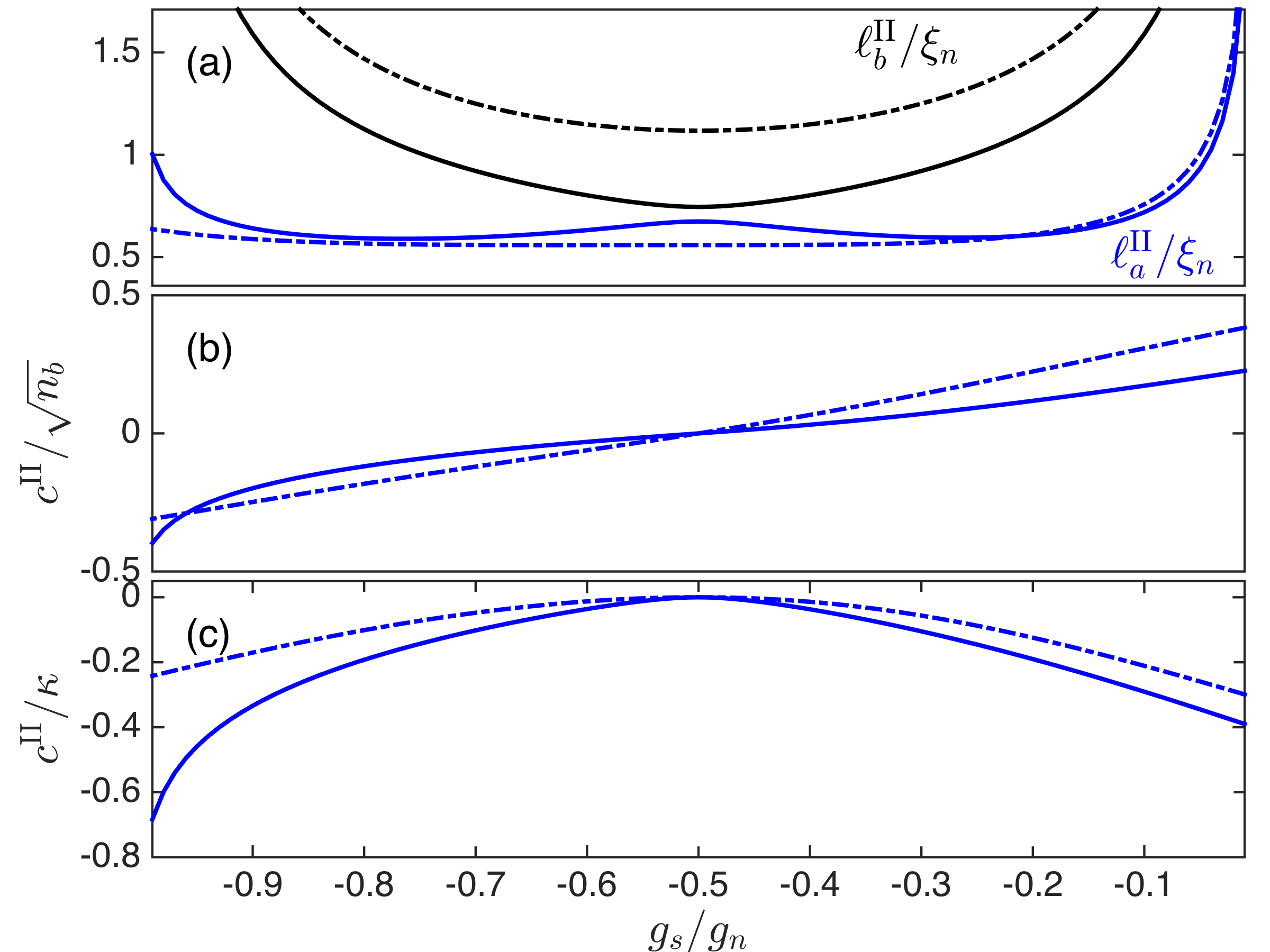}
	\caption{Key parameters in the ansatz of type-II FDSs [Eqs.~\eqref{type-II1} and \eqref{type-II}] as functions of $g_s$ for $\tilde{q}=0.1$ (solid lines) and $\tilde{q}=0.6$ (dash-dotted lines).  \label{f:parameter} } 
\end{figure}

\begin{figure}[htp] 
	\centering
	\includegraphics[width=0.479\textwidth]{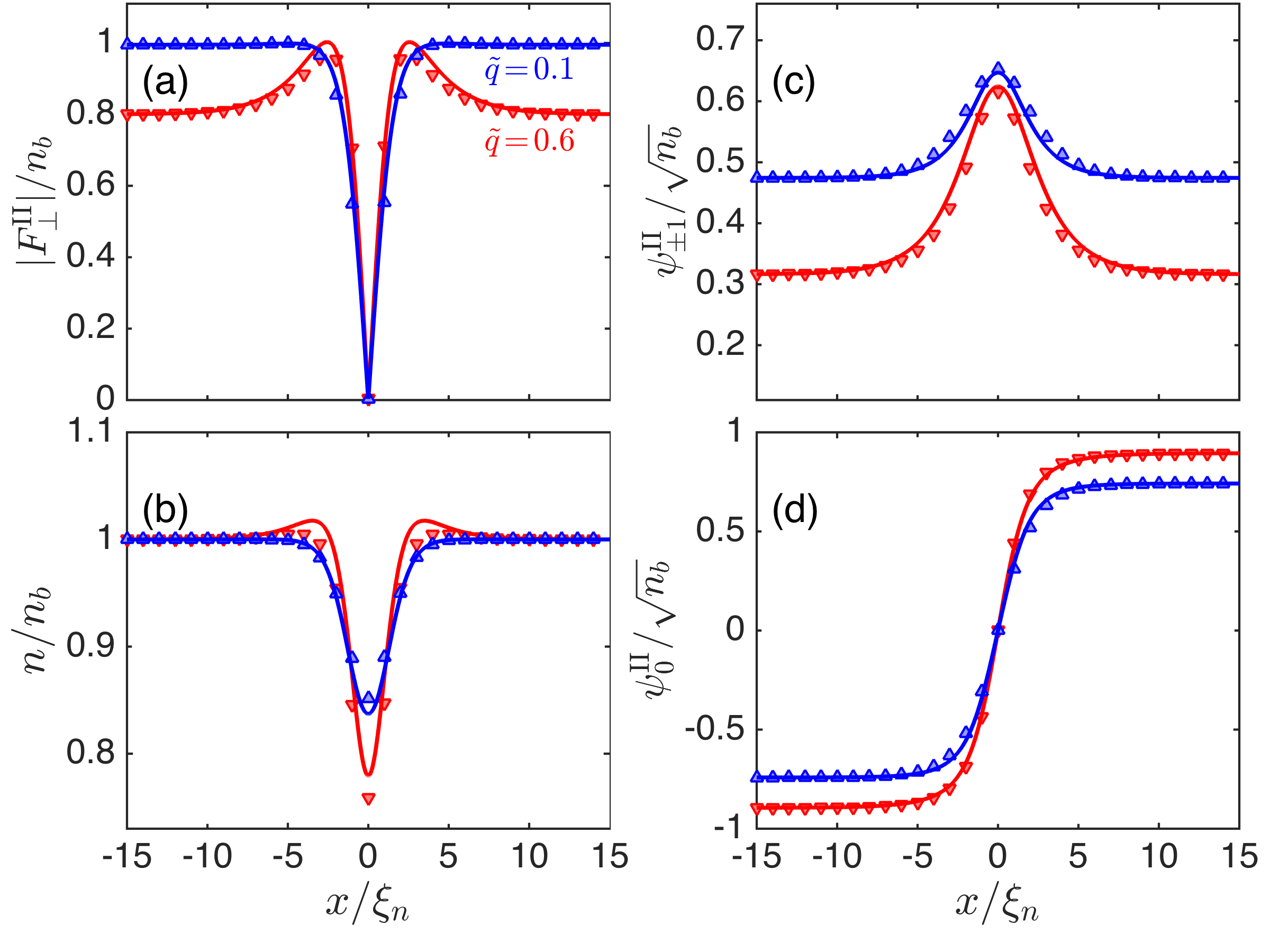}
	\includegraphics[width=0.479\textwidth]{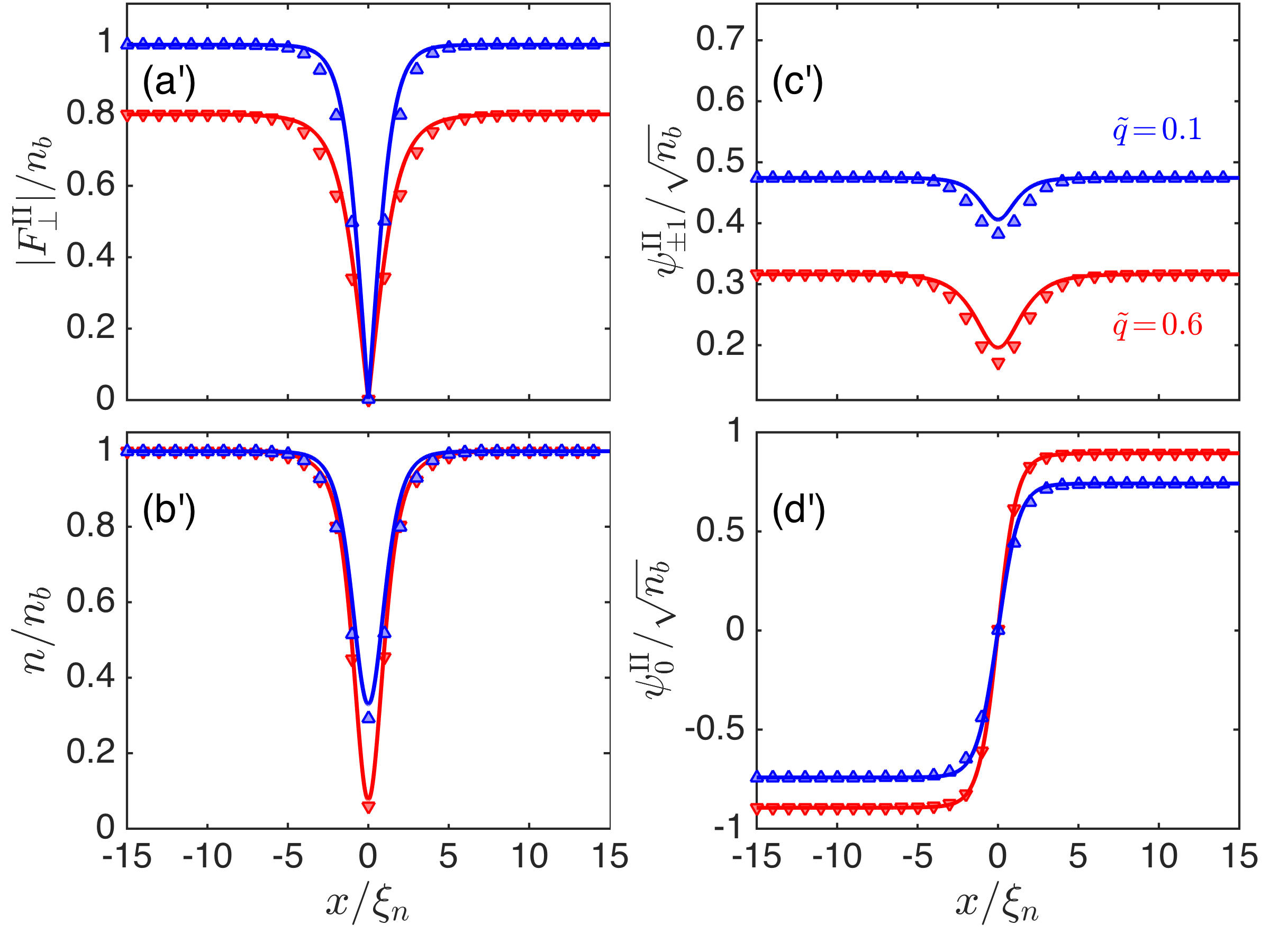}
	\caption{A comparison between analytical predictions of type-II FDSs (solid lines) and numerical results (markers) 
	at $\tilde{q}=0.1$ and $\tilde{q}=0.6$. 	(a), (b),(c) and (d) are for $g_s/g_n=-0.1$ ; (a$'$), (b$'$),  (c$'$) and (d$'$) are for $g_s/g_n=-0.7$.  \label{f:profiletypeII}} 
\end{figure}

At the exactly solvable point  Eqs.~\eqref{SGPE1} and~\eqref{SGPE2} admit  another exact solution -- the type-II soliton given in Eq.~\eqref{typeIIexact}.
Away from the exactly solvable point, the core structure of type-II FDSs exhibits additional complexity which requires two length scales to describe the core widths near and away from the center. 
We hence propose the following ansatz
\bea
\psi^{\rm II}_{0}(x)&&=\sqrt{n^{g}_0} \tanh \left(\frac{x}{2 \ell^{\rm II}_a}\right) \left(\frac{c^{\rm II}/\kappa}{\cosh \left(\frac{x}{\ell^{\rm II}_b}\right)}+1\right),\label{type-II1}\\ 
\psi^{\rm II}_{\pm 1}(x)&&=\frac{c^{\rm II} }{\cosh \left(\frac{x}{\ell^{\rm II}_b}\right)}+\sqrt{n^{g}_{\pm 1}}\label{type-II},
\eea
where we assume $\ell^{\rm II}_b>\ell^{\rm II}_a$. For $x \gg \ell^{\rm I}_{b}$,  $\psi^{\rm II}_{0}(x) \sim \sqrt{n^{g}_0} \left[1+2(c^{\rm II}/\kappa) e^{-x/\ell^{\rm II}_b}\right]$ and  $\psi^{\rm II}_{\pm 1}(x) \sim \sqrt{n^{g}_{\pm1}}+ 2 c^{\rm II}  e^{-x/\ell^{\rm II}_b}$. Solving Eqs.~\eqref{SGPE1} and ~\eqref{SGPE2} asymptotically, we obtain that $\ell^{\rm II}_b$ has to be a root of Eq.~\eqref{widthtypeI}
and 
\bea
\kappa=\frac{g_n q-2 g_s \sqrt{n_b^2 (g_n+2 g_s)^2-\frac{q^2 (g_n+g_s)}{g_s}}}{2 g_s (g_n+2 g_s) \sqrt{\frac{2 q}{g_s}+4 n_b}}.
\eea
If the core was characterized by a single length scale, the value of $\ell^{\rm II}_{b}$ could  be easily set to be $y_{+}$, as it gives rise to the right value at the exactly solvable point, namely  $y_{+}|_{g_s=-g_n/2}=\ell^{\rm II}_{\rm ex}$. However, when the other length scale $\ell^{\rm II}_a$ is introduced, a different scenario becomes possible to connect to the exactly solvable point, i.e., $c^{\rm II} \rightarrow 0$ and $\ell^{\rm II}_a \rightarrow \ell^{\rm II}_{\rm ex}$ as $g_s \rightarrow -g_n/2$.  
The appropriate choice for $\ell_b^{\rm II}$ to provide a consistent solution is
\bea
\ell^{\rm II}_{b}=y_{-}=\ell^{\rm I}.
\eea  
Near the core $x\sim 0$,  solving Eqs.~\eqref{SGPE1} and ~\eqref{SGPE2} to leading order, we obtain
\bea
\ell^{\rm II}_a=\sqrt{\frac{ \hbar ^2}{2M \left(2\mu-{\cal D}\right)}}
\eea
with
\bea
{\cal D}=4(g_n+2 g_s) (c^{\rm II}+\sqrt{n_{\pm 1}})^2 +\frac{3 c^{\rm II} \hbar^2}{(\ell^{\rm I})^2 M (c^{\rm II}+\kappa)},
\eea
where $c^{\rm II}$ is determined by the real solution of
\bea
2 g_n \left(c^{\rm II} +\sqrt{n_{\pm 1}}\right)^3+\frac{c^{\rm II} \hbar ^2}{2 M (\ell^{\rm I})^2}+(q-\mu ) \left(c^{\rm II} +\sqrt{n_{\pm 1}}\right)=0. \nn\\
\eea
It is easy to check that as $g_s\rightarrow -g_n/2$, $c^{\rm II}\rightarrow 0$, ${\cal D}\rightarrow 0$ and $\ell^{\rm II}_a\rightarrow \ell^{\rm II}_{\rm ex}$. Consistently, the two scales we obtained indeed satisfy the working assumption $\ell^{\rm II}_b>\ell^{\rm II}_a$ [Fig. \ref{f:parameter}(a)].
The sign change of $c^{\rm II}$ across $g_s=-g_n/2$ (Fig.~\ref{f:parameter}), yields a hump or a dip in $\psi_{\pm 1}$ (Fig.~\ref{f:profiletypeII}). Note that the other working assumption (i.e.~$\ell^{\rm II}_b<\ell^{\rm II}_a$) does not lead to a consistent solution. A comparison of the two-scale ansatz Eqs.~\eqref{type-II1} and \eqref{type-II} to numerical solutions of the GPEs is presented in Fig.~\ref{f:profiletypeII}. These results validate that this ansatz effectively captures the core structure of type-II FDSs over a wide parameter regime. It is worthwhile to mention that at $q\neq 0$ stationary type-I and type-II FDSs are orthogonal, namely, the overlap $\int dx\, (\psi^{\rm I})^{\dag} \psi^{\rm II}=0$. For propagating FDSs, the overlap between the two types does not vanish and reaches the maximum value at the speed limit~\cite{YuBlairmovingMDW}.

\subsection{Excitation energies}
\label{excitationenergy}
\begin{figure}[htp] 
	\centering
	\includegraphics[width=0.46\textwidth]{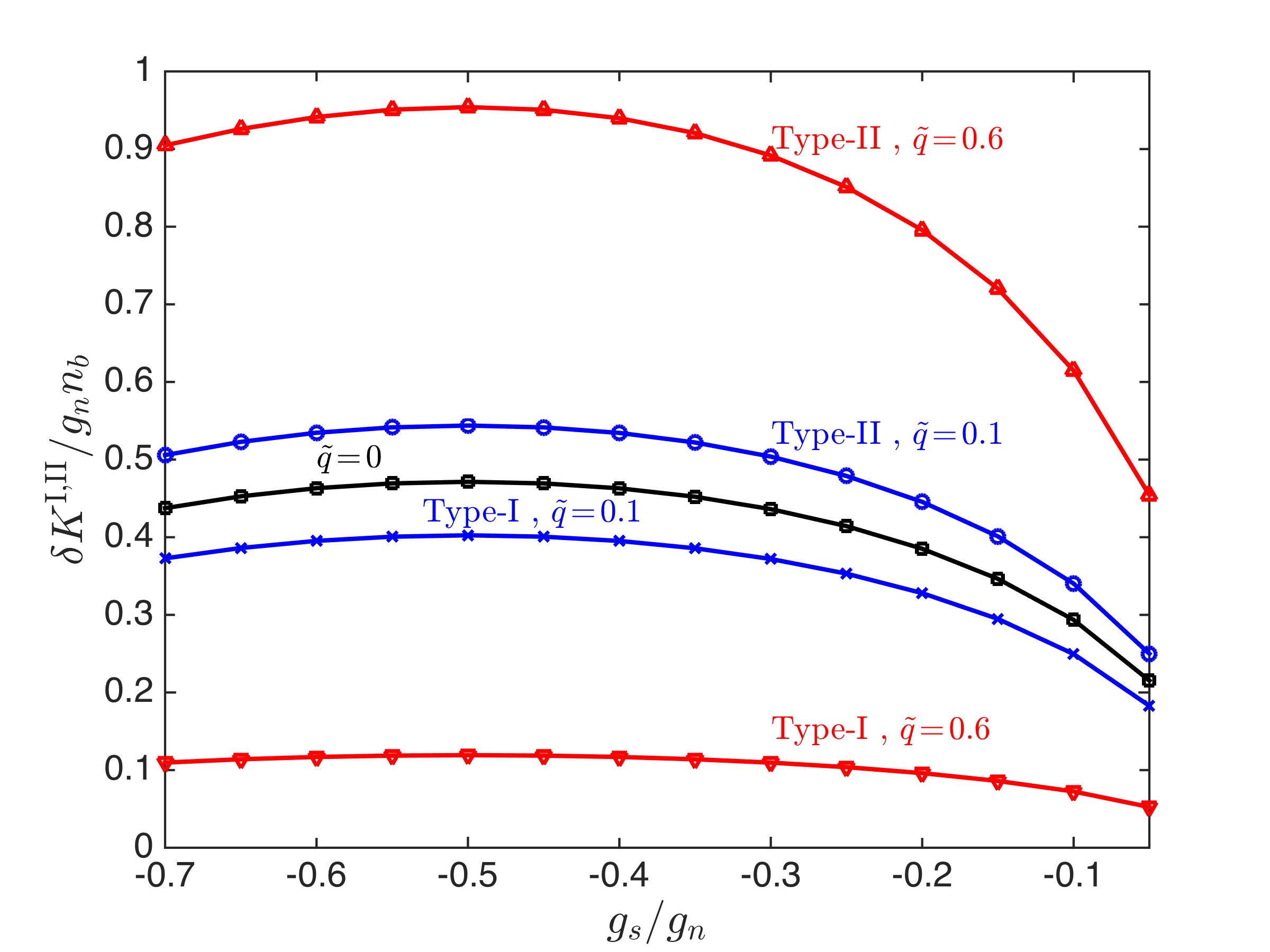}
	\caption{Numerical results of grand canonical energies for type-I and type-II FDSs at $\tilde {q}=0$, $\tilde {q}=0.1 $ and $\tilde{q}=0.6$ (markers). Solid lines are guides to the eye. The soliton energy exhibits a non-monotonic behavior as a function of $g_s/g_n$ and the turning point occurs at the exactly solvable point $g_s/g_n=-0.5$.  At $q=0$ (black squares), type-I and type-II FDSs are degenerate as for this case their states only differ  by a $\textrm{SO}(3)$ spin rotation and a $\textrm{U}(1)$ gauge transformation~\cite{MDWYuBlair}. \label{f:energy}} 
\end{figure} 
At $q=0$, type-I and type-II FDSs are degenerate and are connected by a $\textrm{SO}(3)$ spin-rotation~\cite{MDWYuBlair}. At finite $q$ the magnetic field  breaks the $\textrm{SO}(3)$ rotational symmetry and the degeneracy is lifted. We can characterize the degeneracy breaking using the soliton energy [see Eq.~(\ref{solitonenergy})] $\delta K^{\rm I, II}=K^{\rm I,II}-K_g$,
where $K^{\rm I, II}=K[\psi^{\rm I,II}]$. We find that $\delta K^{\rm I}\le \delta K^{\rm II}$ and the equality is reached only when $q=0$ (see Fig.~\ref{f:energy}).

\subsection{Spin-singlet amplitudes}
So far we have been focused on analyzing the structure of  the order parameter $\mathbf{F}$ and the mass superfluid density $n$ which are the most relevant quantities to characterize FDSs. However, a complete description requires additional information.  In general, 
the set $\{n, \mathbf{F}, \alpha, \delta F^2_z\}$ provides a complete description of a spin-1 BEC, where 
\bea
\alpha \equiv \psi^2_0-2\psi_{+1} \psi_{-1}
\eea
is the spin singlet amplitude and $\delta F^2_z \equiv \psi^{\dag} S^2_z\psi=n_{+1}+n_{-1}$ couples to the quadratic Zeeman energy $q$ [see Eq.~\eqref{Hamioltonian}]. The spin-singlet amplitude $\alpha$ is  $\textrm{SO}(2)$ invariant and 
satisfies 
\bea
|\mathbf{F}|^2+|\alpha|^2=n^2.
\label{singletrelation}
\eea

Let us consider variations that keep $\mathbf{F}$, $n$ and $\delta F^2_z$ unchanged. Such variations appear in the phase of $\alpha$. In the following  we show that this particular phase change is nothing but the $\textrm{U}(1)$ phase variation which is directly related to the mass current
\bea
\mathbf{J}=\frac{\hbar}{2M i}\left(\psi^{\dag}\cdot \nabla \psi-\rm{H.c.}\right)
\eea
satisfying the continuity equation 
\bea
\frac{\partial n}{\partial t}+\nabla \cdot \mathbf{J}=0.
\label{numbercontinuity}
\eea 
For such variations, the deformed wavefunctions must have the form $\widetilde{\psi}_i=\psi_i e^{i\delta \theta_i(\mathbf{r})}$ and the corresponding transverse magnetization reads   
\bea
\hspace{-0.3cm}\widetilde{F}_{\perp}=\sqrt{2}\left(\psi_0\psi^{*}_{+1}e^{i[\delta \theta_0(\mathbf{r})-\delta\theta_{+1}(\mathbf{r})]}+\psi^*_0\psi_{-1}e^{i[\delta \theta_{-1}(\mathbf{r})-\delta\theta_{0}(\mathbf{r})]}\right).
\eea
In order to keep invariant upon a global $\textrm{SO}(2)$ spin rotation, namely $\widetilde{F}_{\perp}=e^{i\phi} F_{\perp}$, the condition 
\bea
\delta \theta_0(\mathbf{r})-\delta\theta_{+1}(\mathbf{r})=\delta \theta_{-1}(\mathbf{r})-\delta\theta_{0}(\mathbf{r})=\phi
\label{phasefluctuation}
\eea 
must be satisfied, where $\phi$ is an arbitrary constant phase.  Phase fluctuations satisfying Eq.~\eqref{phasefluctuation} are captured by the spin-singlet, as  
$\widetilde{\alpha}= e^{2i \delta \theta_0(\mathbf{r})}\alpha=e^{2i [\delta \theta_{+1}(\mathbf{r})+\phi]}\alpha=e^{2i [\delta \theta_{-1}(\mathbf{r})-\phi]}\alpha$. It is also easy to see that $\delta \theta_0(\mathbf{r})$ or $\delta \theta_{\pm1}(\mathbf{r})$ is an overall $\textrm{U}(1)$ phase variation and 
$\widetilde{\mathbf{J}}=\mathbf{J}+n \nabla \delta \theta_0(\mathbf{r})=\mathbf{J}+n \nabla \delta \theta_{\pm1}(\mathbf{r})$.
It is clear that a general phase variation of $\alpha$ does not necessarily  have to be an overall $\textrm{U}(1)$ phase change.

\begin{figure}[htp] 
	\centering
	\includegraphics[width=0.246\textwidth]{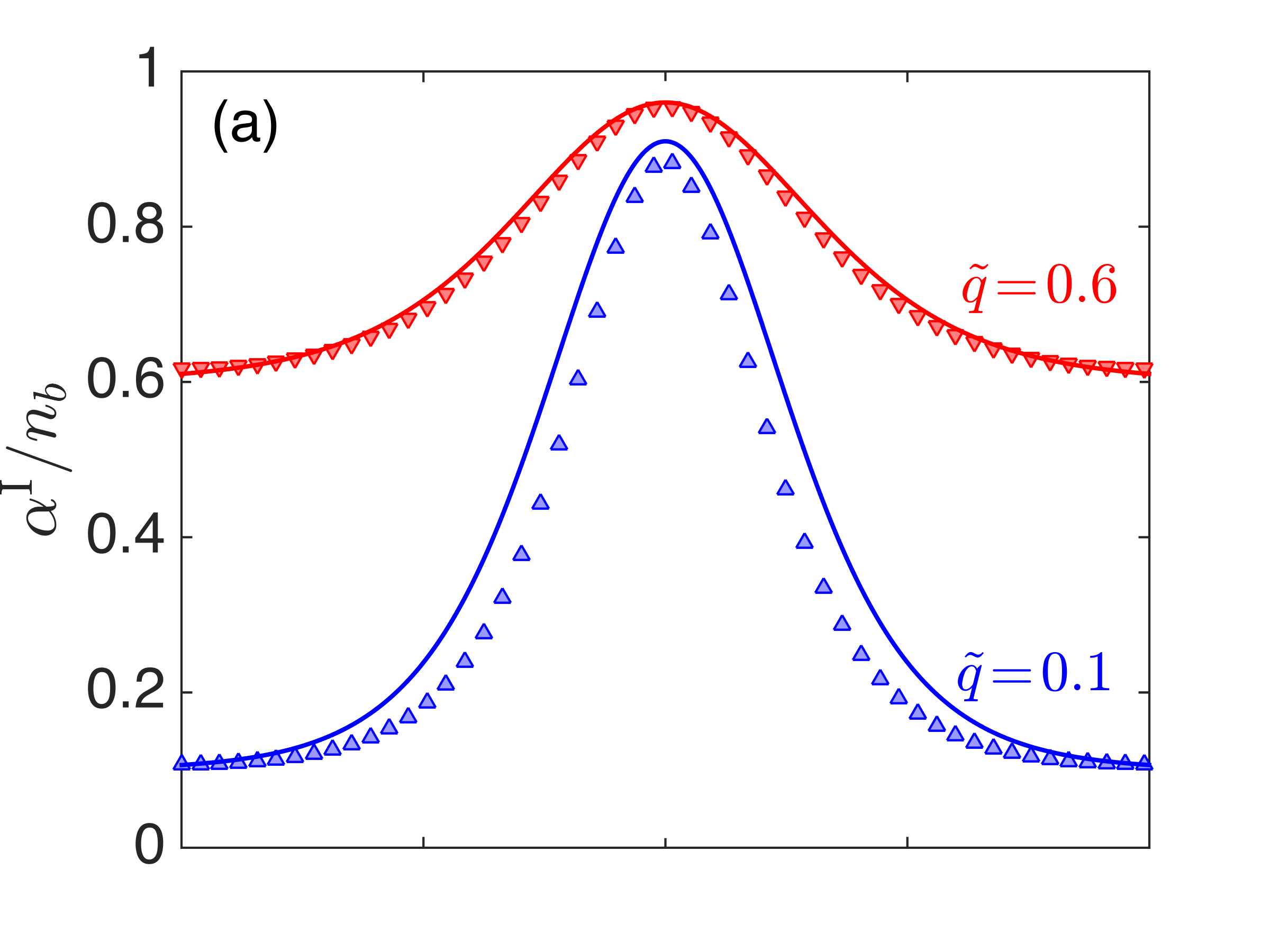}
	\vspace*{-0.336cm} \hspace*{-0.37cm}
	\includegraphics[width=0.246\textwidth]{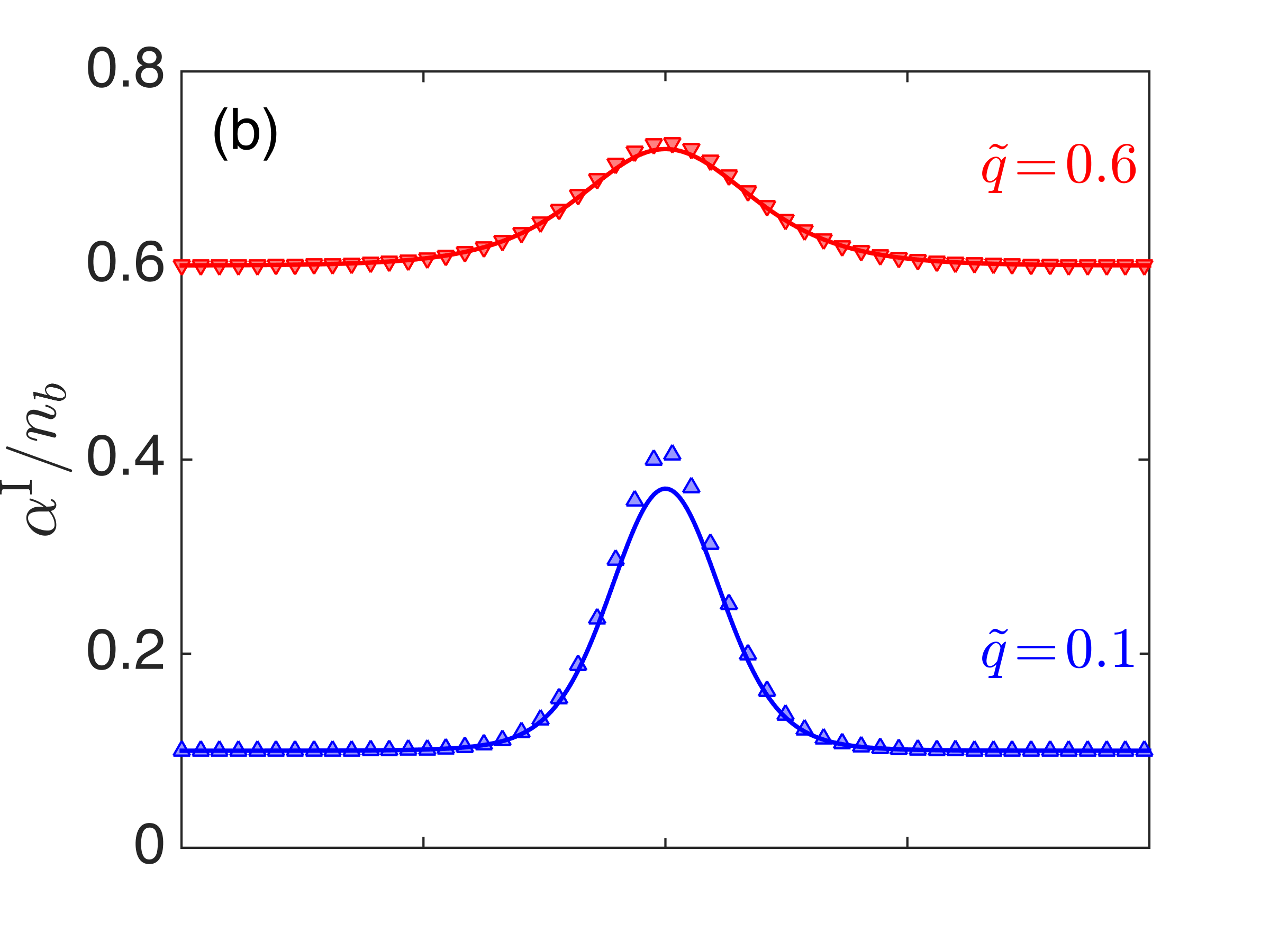}
	\includegraphics[width=0.246\textwidth]{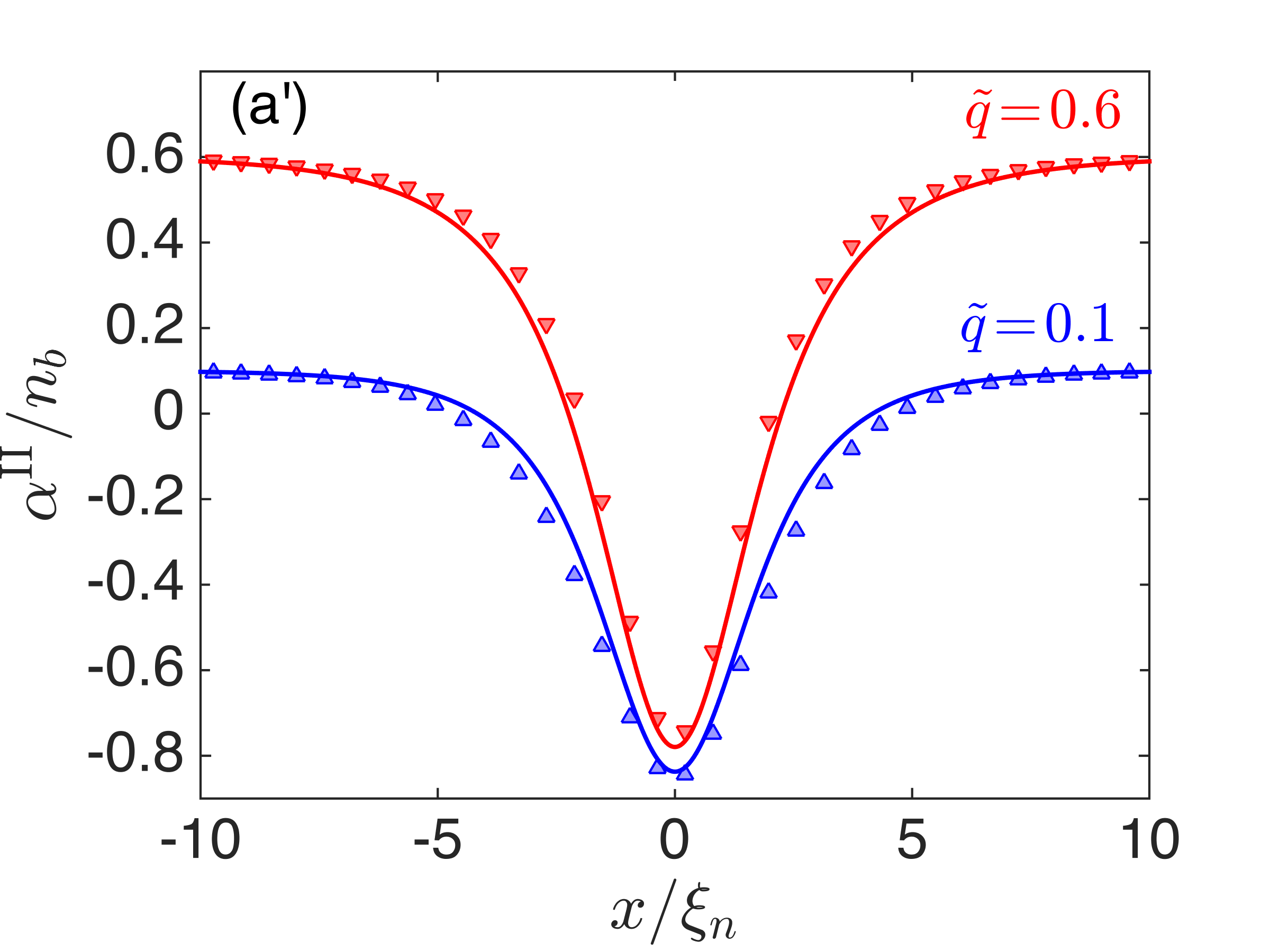}
	\hspace*{-0.37cm}
	\includegraphics[width=0.246\textwidth]{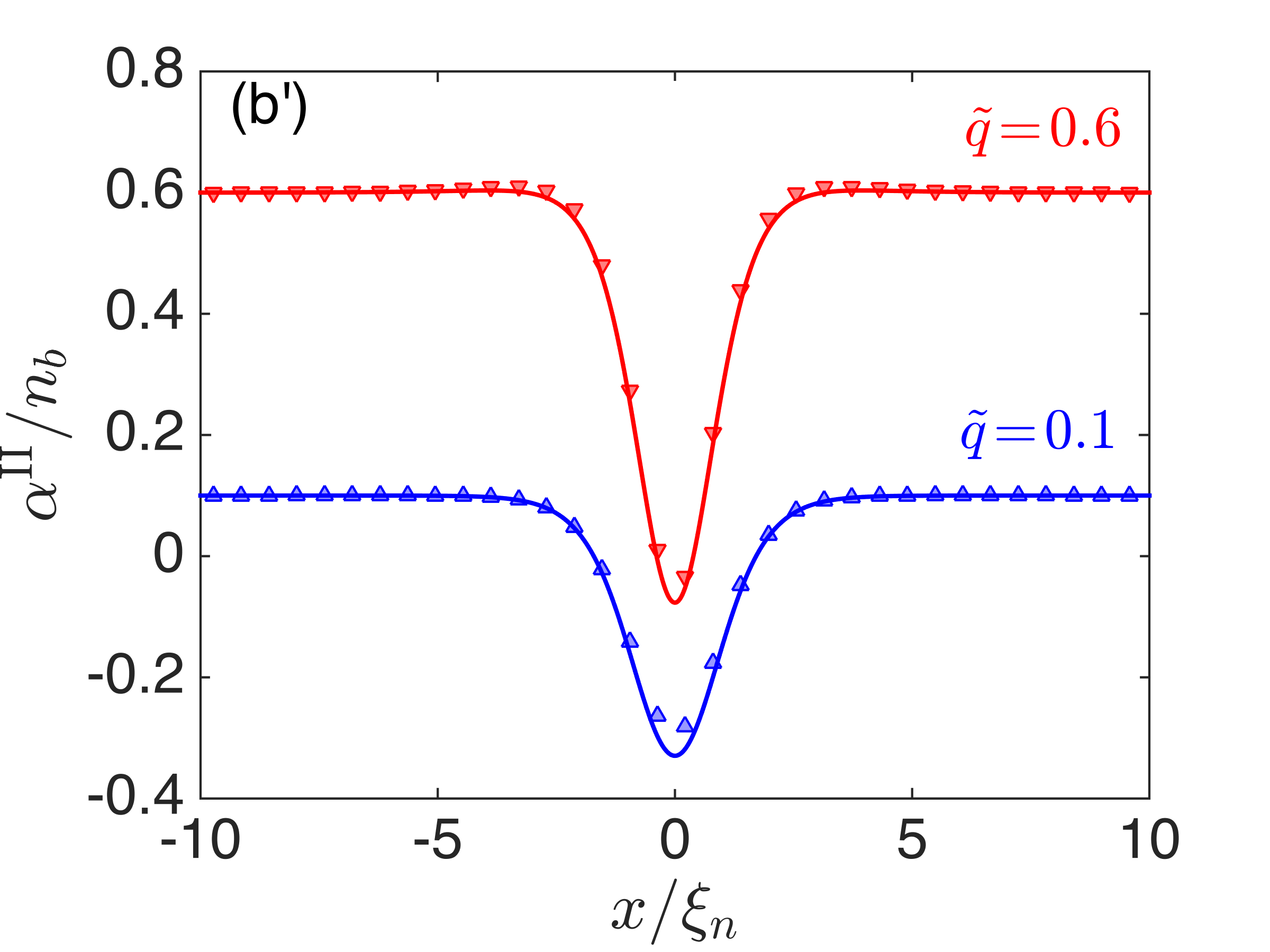}
	\caption{Analytical predictions (solid lines) and numerical results (markers) of the spin-singlet amplitude $\alpha$ for type-I [(a),(b)] and type-II [(a$'$),(b$'$)] FDSs at $\tilde{q}=0.1$ and $\tilde{q}=0.6$. (a) and (a$'$) are for   $g_s/g_n=-0.1$; (b) and  (b$'$) are for $g_s/g_n=-0.7$.     \label{f:spinsinglet} } 
\end{figure}

For FDSs, comparing to $n$, $\mathbf{F}$ and $\delta F^2_z$, the spin-singlet amplitude $\alpha$ is a more appealing quantity to distinguish type-I from type-II FDSs as it shows a hump or a dip depending on the type (Fig.~\ref{f:spinsinglet}).  The measurement of $\alpha$ is in the scope of the current spin-1 BECs experiments~\cite{Kunkel2019}.

\section{Nematic structure}
It is sometimes useful to formulate the spin-1 BEC dynamics in terms of 
physical observables. The hydrodynamic formulation of the GPEs Eqs.~\eqref{spin-1GPE} and \eqref{spin-1bGPE} serves this purpose and provides a complete description of the condensate dynamics~\cite{currents}. Such a description naturally involves nematic tensors and the corresponding currents. Although $n$, $\mathbf{F}$, $\alpha$ and $\delta F^2_z$ together also fully describe a spin-1 BEC, it is not anticipated that the construction of self-contained closed dynamical equations with these variables is achievable. In this section, we explore nematic structure of FDSs.

\subsection{General formulation}
The magnetization continuity equation reads
\bea
\frac{\partial F_i}{\partial t}+\nabla \cdot J^{F}_i=K_{iz},
\label{MC}
\eea 
where the spin current density
\bea
J^{F}_i=\frac{\hbar}{2M i} \left(\psi^{\dag} S_i \nabla \psi-\nabla \psi^{\dag} S_i \psi \right),
\eea 
and the source term (or the internal current density)
\bea
K_{iz}=\frac{2 q}{\hbar}\sum_{k}\epsilon_{izk}N_{zk}.
\eea
Explicitly, the components are $K_{xz}=- (2 q/\hbar) N_{zy}$, $K_{yz}=(2 q/\hbar) N_{zx}$ and $K_{zz}=0$, where $N_{ij}$ is the nematic (or quadrupolar) tensor density
\bea
N_{ij}=\psi^{\dag}\hat{N}_{ij}\psi
\eea
with $\hat{N}_{ij}=(S_iS_j+S_jS_i)/2$ and $i,j\in\{x,y,z\}$. The nematic tensor serves as an order parameter if $F_i=0$~\cite{nematic,symes2017} and describes spin fluctuations in the ferromagnetic phase. It is easy to recognize that $N_{zz}=\delta F^2_z$.

The nematic continuity equation reads
\bea
\frac{\partial N_{ij}}{\partial t}+\nabla \cdot J^{N}_{ij}=W_{ij}+Q_{ijz},
\label{NC}
\eea 
where the nematic current density
\bea
J^{N}_{ij}=\frac{\hbar}{2Mi} \left(\psi^{\dag} \hat{N}_{ij} \nabla \psi-\nabla \psi^{\dag} \hat{N}_{ij} \psi\right),
\eea 
and the source terms (or the internal current densities)
\bea
W_{ij}&&=\frac{g_s}{\hbar} \sum_{l,k}F_{l} \left(\epsilon_{i l k} N_{k j}+\epsilon_{j l k} N_{k i}\right), \\
Q_{ijz}&&=\frac{q}{2\hbar}\sum_{k} F_{k} \left(\epsilon_{izk}\delta_{zj} + \epsilon_{jzk}\delta_{zi}\right).
\eea 
Since $\Tr{N}=\sum_i{N_{ii}}=n$, the total density continuity equation~\eqref{numbercontinuity} can be obtained by taking the trace of Eq.~\eqref{NC}. It is easy to see that $\Tr{W}=0$ and $Q_{iiz}=0$. Also
$Q_{xzz}=Q_{zxz}=-(q/2\hbar)F_y$ and $Q_{yzz}=Q_{zyz}= (q/2\hbar) F_x$,  and all the other components are zero.  Hence $Q_{ijz}$ does not contain new information. 
The continuity equations \eqref{numbercontinuity}, \eqref{MC}, \eqref{NC} and the equation of motion of the  mass current $\mathbf{J}$ (Euler equation) together provide a complete description of spin-1 BECs dynamics~\cite{currents}.   

Except the total number density $n$ and the total number  current $\mathbf{J}$, the expressions of other densities and currents vary when applying spin-rotations. Hence it is useful to construct rotationally invariant quantities to reveal intrinsic currents.   Here we find that the densities  
\bea
F\equiv \left[\sum_i \left(F_i\right)^2\right]^{1/2}, \quad {\cal N} \equiv \left[\sum_{ij} \left(N_{ij}\right)^2\right]^{1/2}
\eea
and the current densities 
\bea
{\cal J}_F \equiv && \left[\sum_{i} \left(J^{F}_{i}\right)^2\right]^{1/2},\nn\\
{\cal J}_N \equiv&& \left[\sum_{ij} \left(J^{N}_{ij}\right)^2\right]^{1/2},\nn\\
{\cal W} \equiv &&\left[\sum_{ij} \left(W_{ij}\right)^2\right]^{1/2} 
\eea 
are invariant under $\textrm{SO}(3)$ spin-rotations.
The source terms 
\bea
{\cal Q}\equiv \left[\sum_{ij} \left(Q_{ijz}\right)^2\right]^{1/2},\quad {\cal K} \equiv \left[\sum_{i} \left(K_{iz}\right)^2\right]^{1/2}
\eea
only appear in the equation of motion when $q\neq 0$, and are $\textrm{SO}(2)$ rotationally invariant.

\subsection{Nematic structure of FDSs}
\begin{figure}[htp] 
	\centering
	\includegraphics[width=0.246\textwidth]{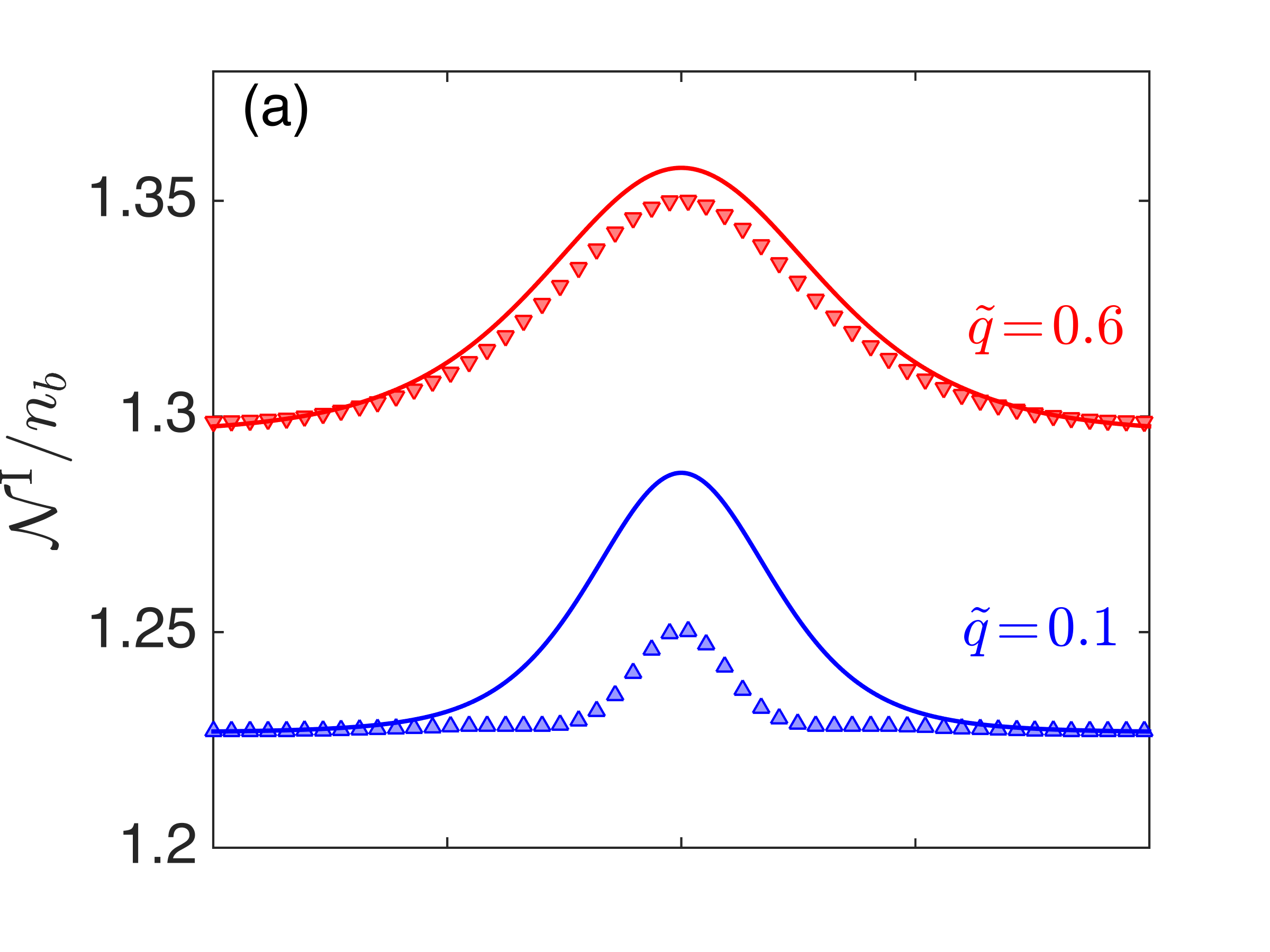}
		\vspace*{-0.336cm} \hspace*{-0.37cm}
	\includegraphics[width=0.246\textwidth]{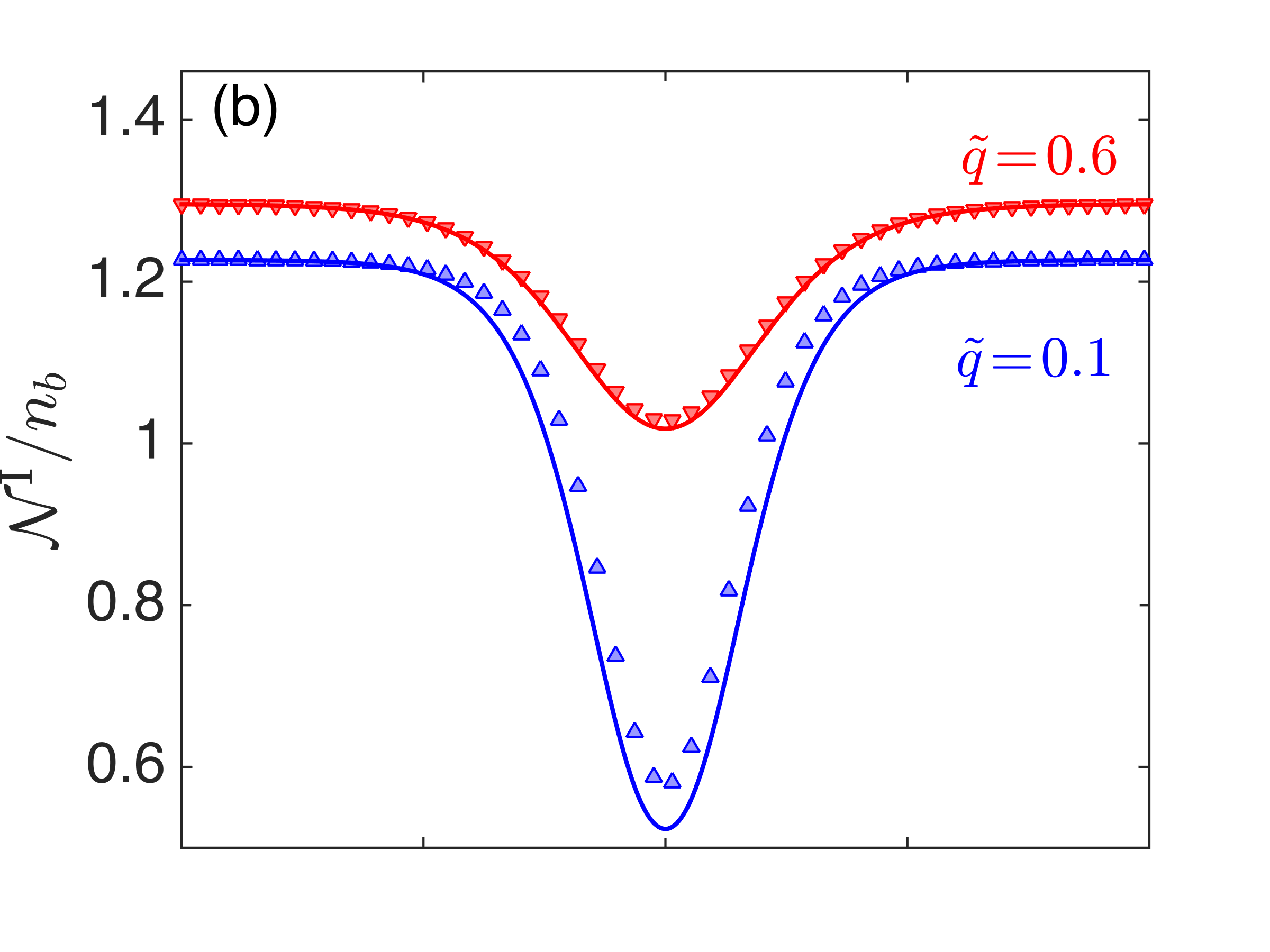}
	\includegraphics[width=0.246\textwidth]{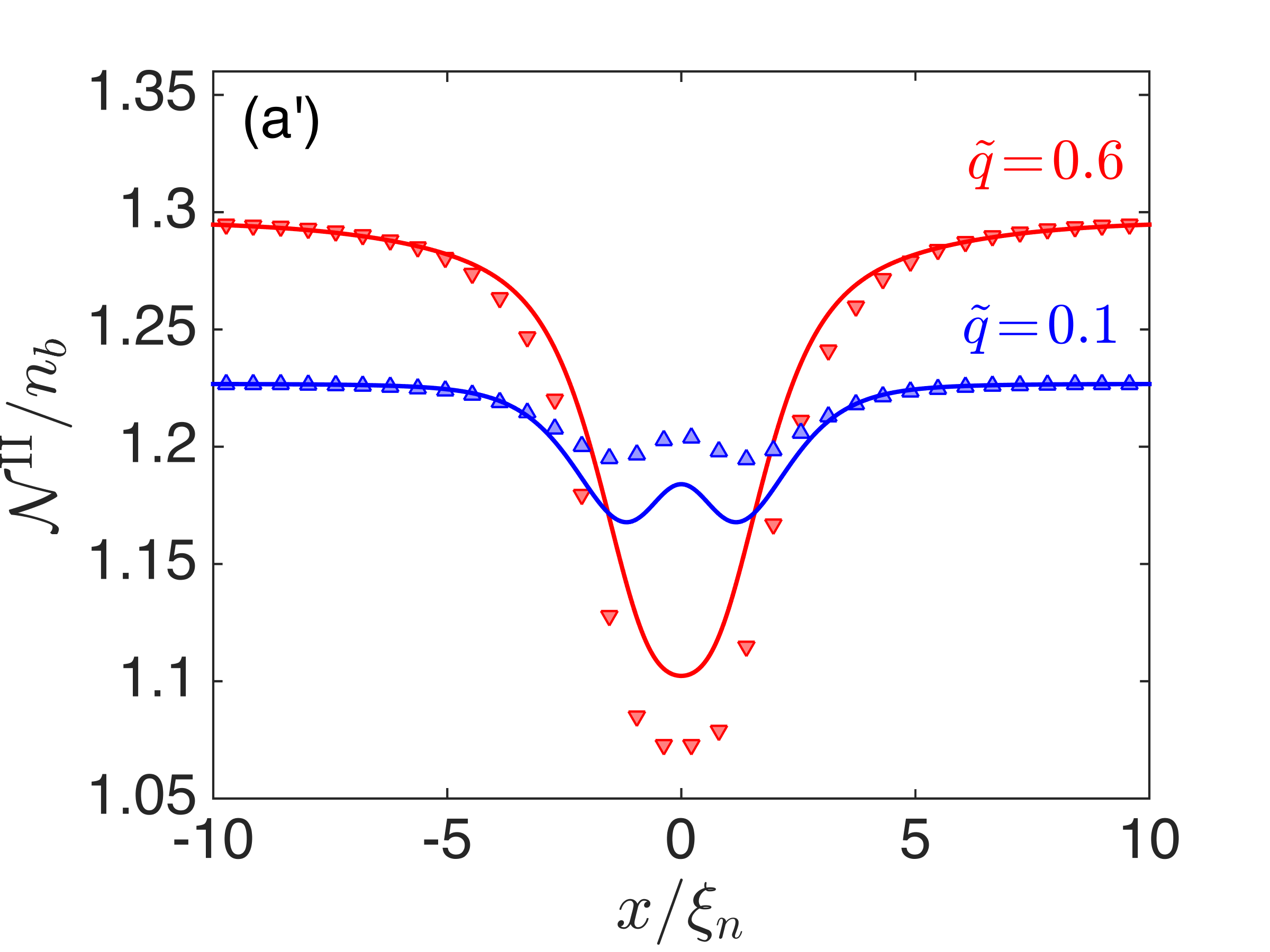}
	\hspace*{-0.37cm}
	\includegraphics[width=0.246\textwidth]{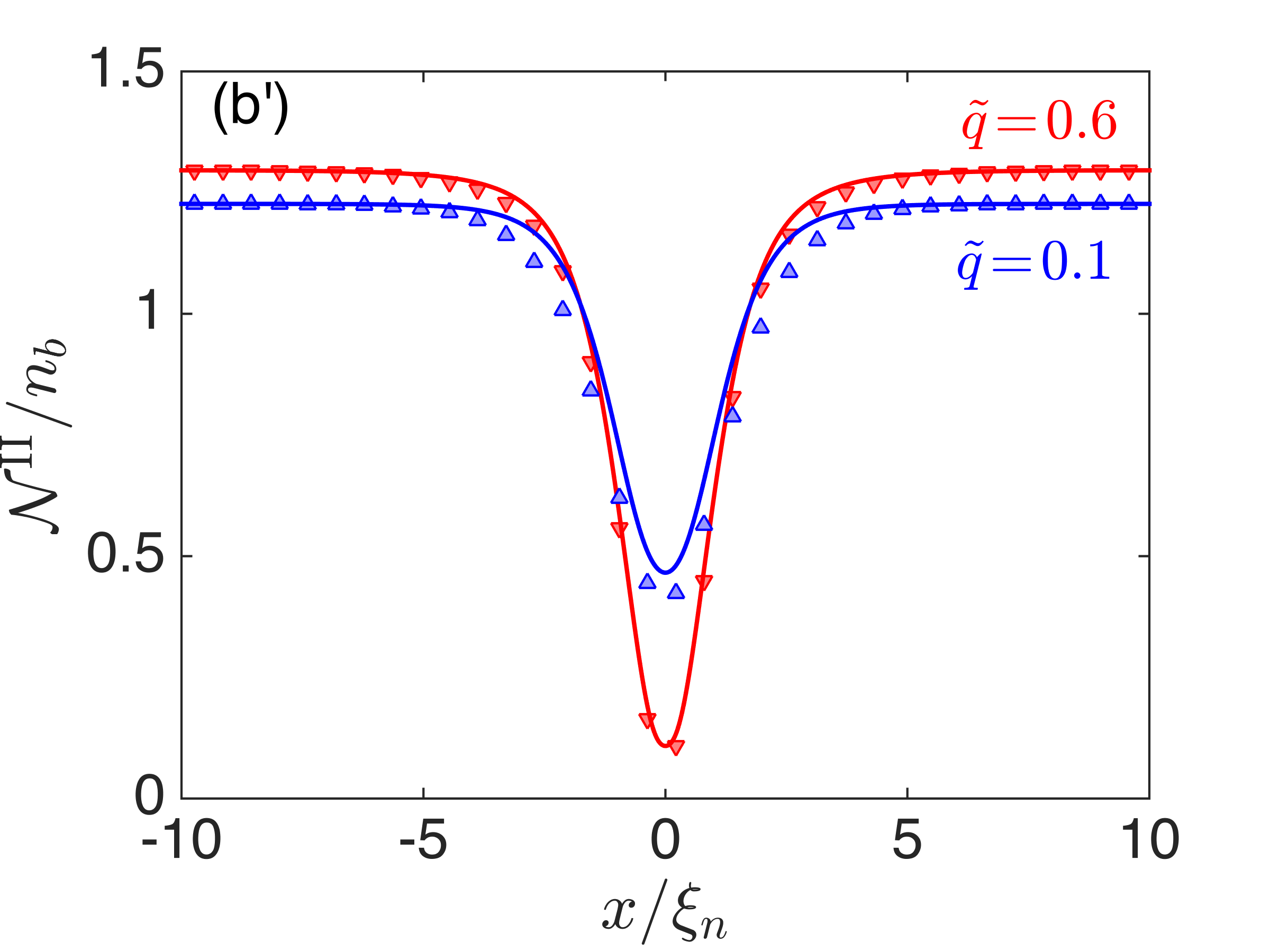}
	\caption{Analytical predictions (solid lines) and numerical results (markers) of the rotationally invariant nematic density ${\cal N}$ for type-I [(a),(b)] and type-II [(a$'$),(b$'$)] FDSs at $\tilde{q}=0.1$ and $\tilde{q}=0.6$. (a) and (a$'$) are for   $g_s/g_n=-0.1$; (b) and  (b$'$) are for $g_s/g_n=-0.7$.   \label{f:nematicprofile}} 
\end{figure}

At the exactly solvable point $g_s=-g_n/2$, the invariant nematic densities read 
\bea
({\cal N}^{\rm I,II})^2 &&=
\frac{1}{4} \left[4 \left(n_b-\frac{g_n n_b\mp q}{g_n \cosh \left(\frac{x}{\ell^{\rm I, II}_q}\right)+g_n}\right)^2 \right. \\
&& \left. +\left(n_b \mp \frac{q}{g_n}\right)^2 \tanh ^4\left(\frac{x}{2 \ell^{\rm I,II}_q}\right)+\left(n_b \pm \frac{q}{g_n}\right)^2\right], \nn
\eea
where the upper (lower) sign in front of $q$ specifies the type-I (type-II) FDS. Away from the exactly solvable point, the ansatzes also describe well the nematic densities in a wide range parameter regime (Fig.\ref{f:nematicprofile}).
 
\begin{figure}[t] 
	\centering
	\includegraphics[width=0.2399\textwidth]{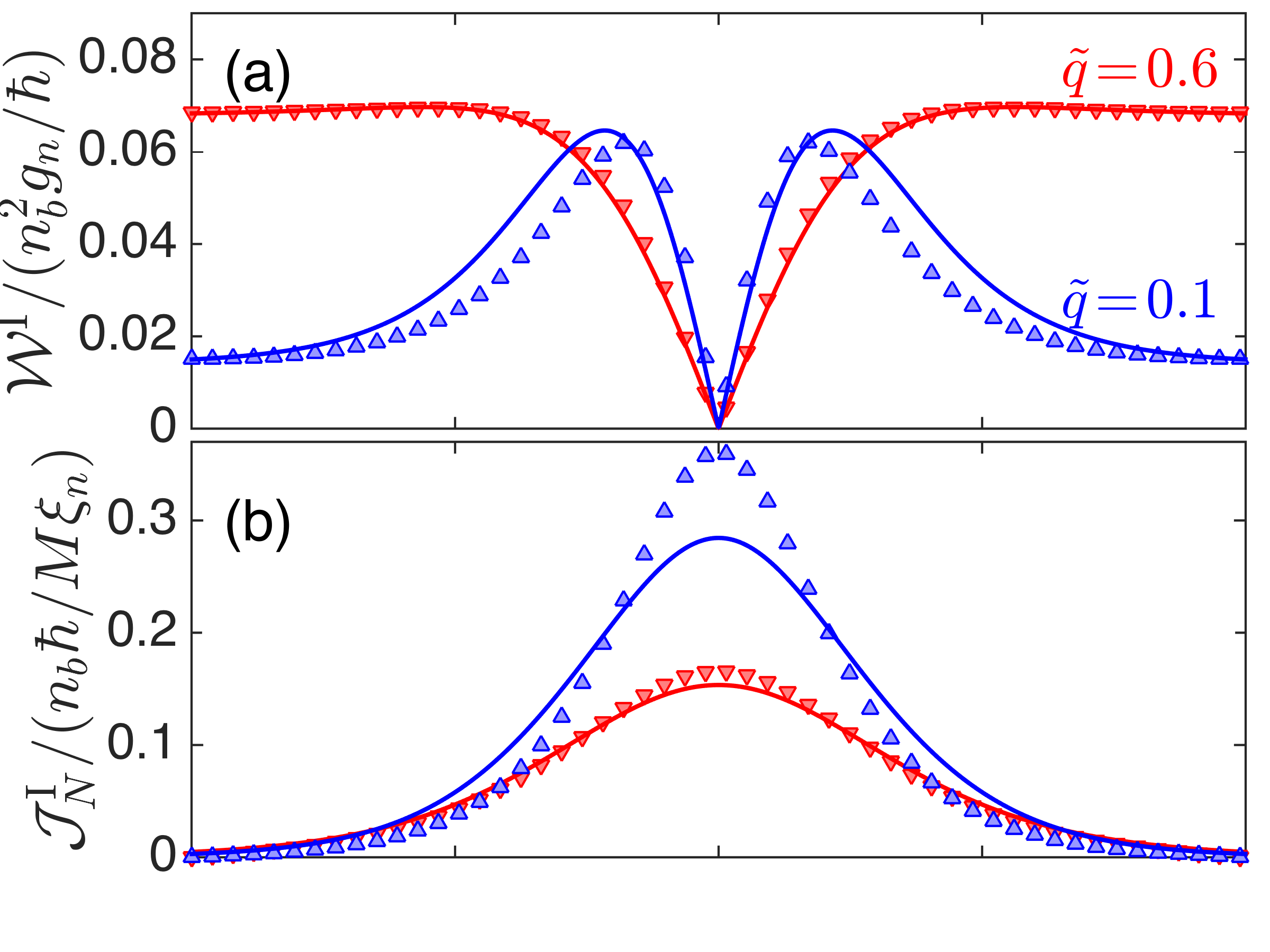}
	\vspace*{-0.28cm} 	\hspace*{-0.15cm}
	\includegraphics[width=0.2399\textwidth]{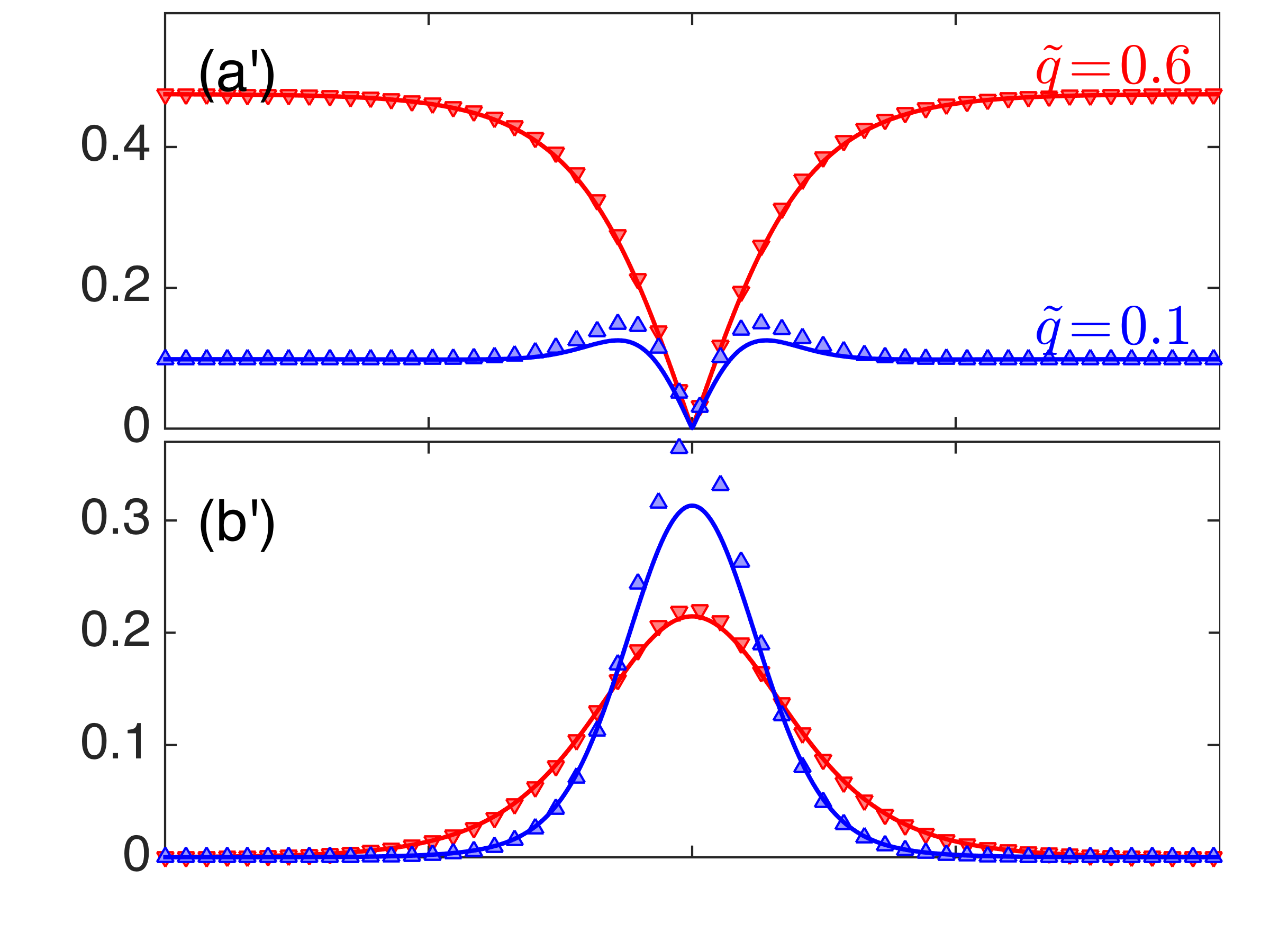}	
	\includegraphics[width=0.2399\textwidth]{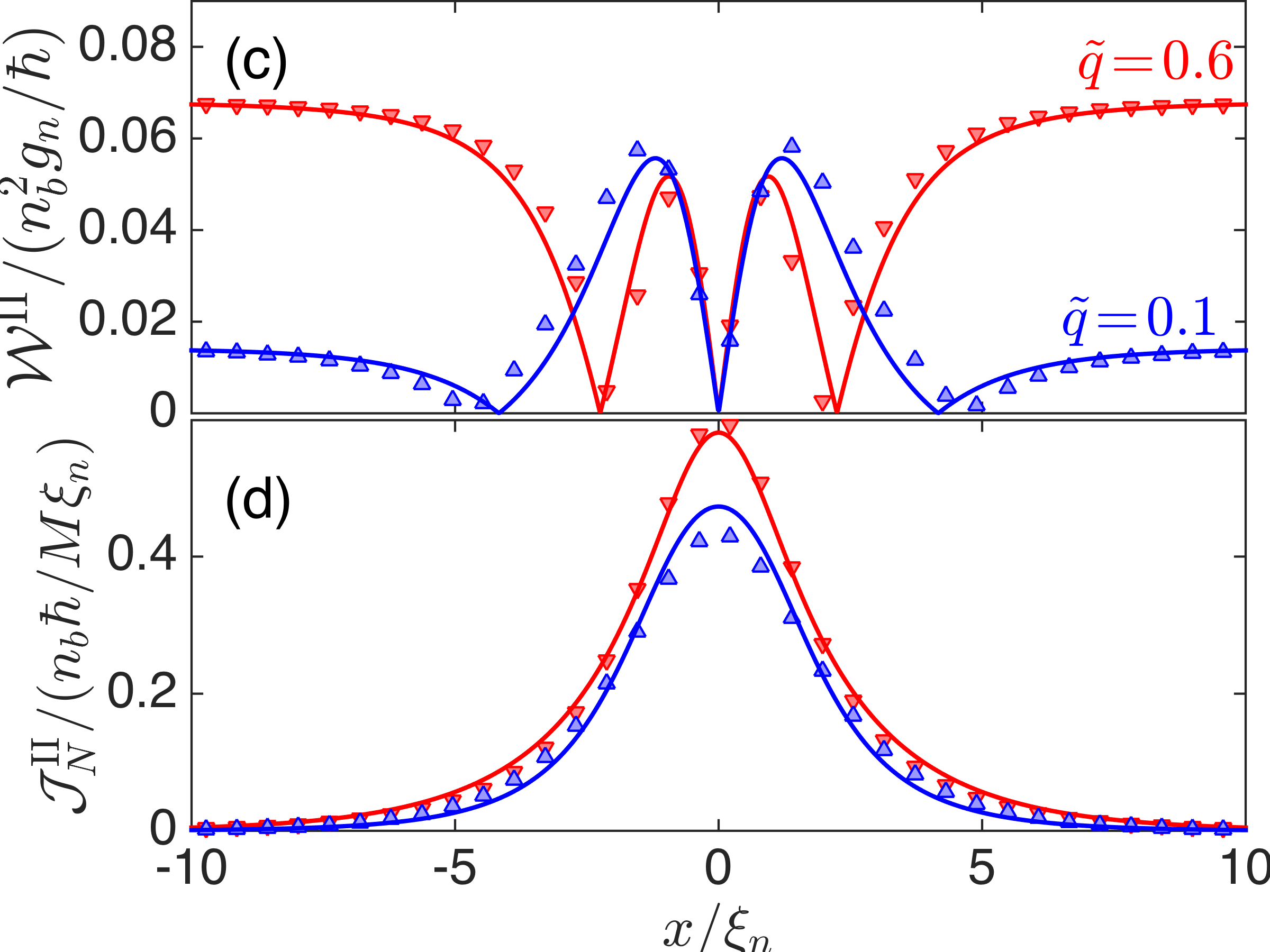}
	\hspace*{-0.15cm}
	\includegraphics[width=0.2399\textwidth]{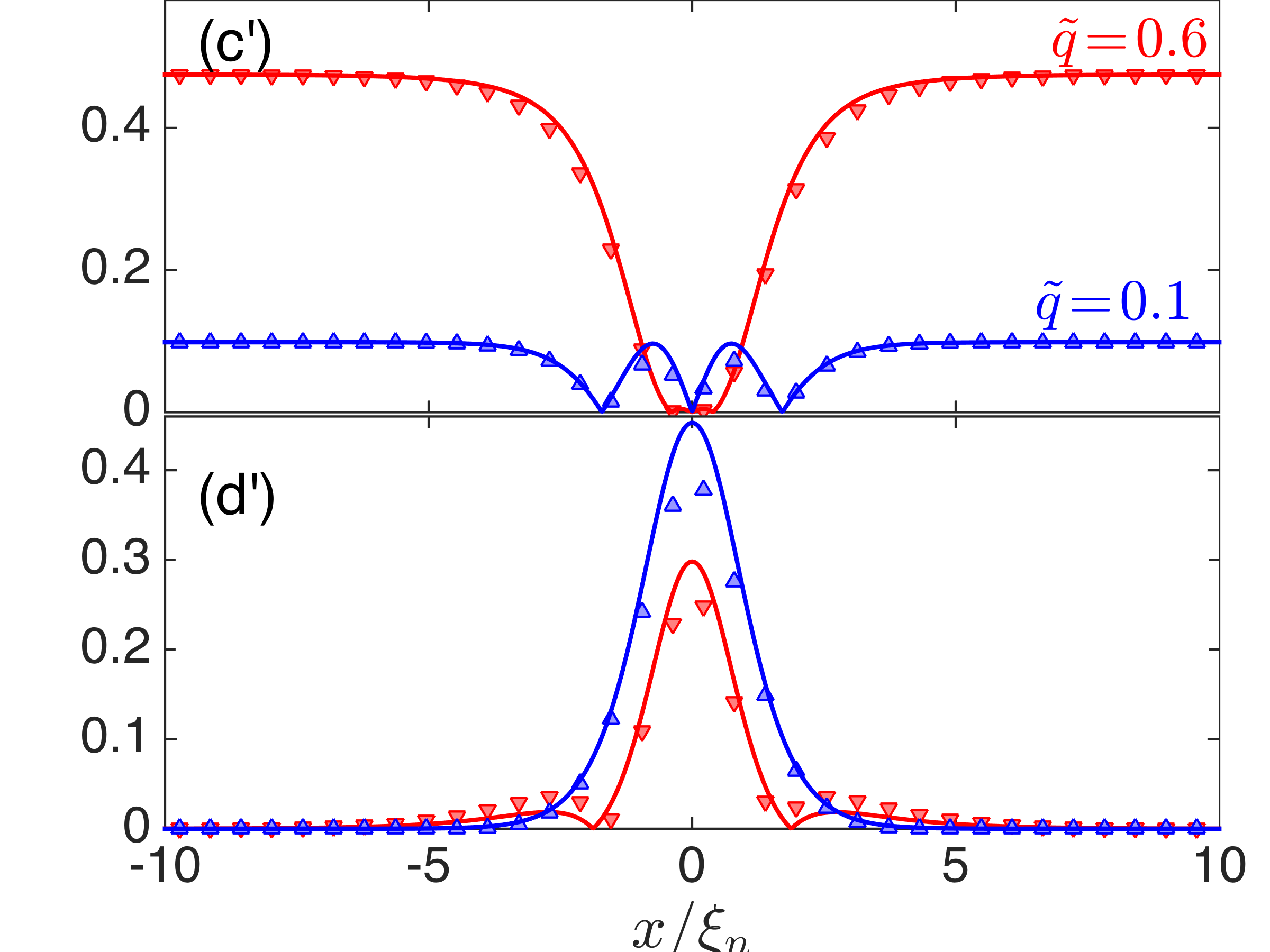}
	\caption{Profiles of invariant internal currents of FDSs. Analytical predictions (solid lines) and numerical results (markers) are shown at $\tilde{q}=0.1$ (blue) and $\tilde{q}=0.6$ (red). (a), (b), (c) and (d) are for  $g_s/g_n=-0.1$; (a$'$), (b$'$), (c$'$) and (d$'$) are for $g_s/g_n=-0.7$. The $y$-axis labels of (a$'$), (b$'$), (c$'$) and (d$'$) are the same as the $y$-axis labels of (a), (b), (c) and (d).  \label{f:profile}} 
\end{figure} 
 
 Figure~\ref{f:profile} shows the profiles of the invariant internal current densities.  At the exactly solvable point, 
\bea
({\cal J}^{\rm I,II}_N)^2&&=\frac{ \hbar^2(g_n^2n_b^2-q^2) }{32 (\ell^{\rm I,II}_q)^2 M^2 g^2_n}\sech^4\left(\frac{x}{2 \ell^{\rm I,II}_q}\right),\\
({\cal W}^{\rm I,II})^2&&=\frac{g^2_n n^2_b-q^2}{8 g_n^2 \hbar ^2}\tanh ^2\left(\frac{x}{2 \ell^{\rm I,II}_q}\right) \times \nn\\
&&\sech^4\left(\frac{x}{2 \ell^{\rm I,II}_q}\right)
\left[g_n n_b \pm q \cosh \left(\frac{x}{\ell^{\rm I,II}_q}\right)\right]^2, 
\eea 
here the plus sign and minus sign in front of $q$ specify type-I and type-II FDSs respectively.

\section{Conclusion and discussion}
We have studied the $\mathbb{Z}_2$ topological defects in the easy-plane phase of a ferromagnetic spin-1 BEC.  We propose analytical ansatzes to describe ferrodark solitons (FDSs) which are the defects in the magnetic order.  The ansatzes reduce to the exact solutions at the exactly solvable point and show good agreement with numerical results over the whole easy-plane phase. The width of type-I FDSs is captured by a single scale while the core structure of type-II FDSs require two scales.  Moreover, exact dark-dark-dark (DDD) vector solitons, which are the $\mathbb{Z}_2$ defects in the mass superfluid order, are also presented. FDSs are expected to play an important role in late time dynamics of 1D quenches~\cite{Prufer2018a,Gasenzer2019} and equilibrium properties at finite temperature~\cite{Maximilian1D2022}.  

Experimental advances now allow plane-confined BECs with flat-bottom traps~\cite{Dalibard2015,Gauthier16} and the measurement of transverse magnetic structures 
by the nondestructive imaging method ~\cite{Higbie2005, Kunkel2019}, opening the possibility for detailed experimental investigations of FDSs. 
The so-called magnetic phase imprinting method, which has been used to experimentally create pairs of magnetic solitons~\cite{MSexp1,MSexp2}, could be a suitable method to create FDSs in a ferromagnetic spin-1 BEC. Very recently, the easy-plane phase of a homogeneous ferromagnetic spin-1 BEC has been prepared in a flat-bottom trapped quasi-one dimensional system~\cite{Maximilian1D2022}. The magnetic phase imprinting method applies circularly polarized light to imprint a phase shift of $\pm \pi$ onto two  $m=\pm 1$ hyperfine components on half the system by using a magnetic shadow~\cite{MSexp1,MSexp2}. This hence realizes a $\pi$ spin rotation of the transverse magnetization of half of the system, seeding the key ingredient of a single FDS.

In this paper we focus on kinks/topological solitons in the easy-plane phase of ferromagnetic spin-1 BECs. A FDS is a $\mathbb{Z}_2$ topological defect which interpolates between oppositely magnetized domains.
From another perspective, it could be useful to view FDSs as members of the soliton family in spin-1 BECs and compare them with other nonlinear waves. Here we give a brief summary of the main focus of recent soliton studies in spin-1 BECs. The list is incomplete, for instance, solitons in systems with optical lattices and dipole-dipole interactions~\cite{PhysRevA.69.053609,PhysRevA.71.053611} are not included.  Also we only consider systems with 
the finite background density, i.e., the positive density-dependent interaction strength.  According to the terminologies, there are mainly two varieties. 
i) \textit{Vector solitons}: 
each component has either dark or bright soliton structure ~\cite{nistazakis2008bright,Busch2001,Liu2009,ThreeComponentSoliton2018,Stefan2020,2021arXiv210907404K} (Ref.~\cite{2021arXiv210907404K} discusses two-dimensional extensions of such excitations, i.e., vortex-bright structures). Various vector solitons in a 1D harmonically trapped spin-1 system have been recently summarized in Ref.~\cite{Katsimiga_2021}.
Vector solitons may or may not correspond to topological structures in the order parameter. There are two interesting examples which are topological. One is the dark-dark-dark vector soliton in ferromagnetic BECs which is the topological defect in the mass superfluid order as discussed in Sec.~\ref{FDS}. 
Another one is the bright-dark-bright vector soliton in an anti-ferromagnetic BEC which manifests itself as a domain wall in the spin director field (the order parameter)~\cite{shin2019}.  ii) \textit{Magnetic solitons}: this soliton was initially introduced in miscible two-component BECs~\cite{MDQu2016} and can be embedded into an anti-ferromagnetic spin-1 BEC in the absence of a magnetic field~\cite{MSexp1,MSexp2,chai2021prr}. One important aspect of this soliton is that it is an analytical solution which is beyond the Manakov limit. The solution is obtained with the constant density approximation which works well for weak spin-dependent coupling $g_s$.  In the context of two-component BECs, it emphasizes that there is no bright or dark soliton structure in the components, while the pseudo magnetization $F_z$ shows a bright soliton structure. When embedding into anti-ferromagnetic spin-1 BECs, its configuration varies under $\textrm{SO}(3)$ spin-rotations~\cite{chai2021prr} as what happens for FDSs at $q\rightarrow 0$ limit~\cite{MDWYuBlair}. The magnetic soliton is not a topological soliton as there is no invariant topological charge associated with it (neither in its original two-component formulation nor in its realizations in an anti-ferromagnetic spin-1 BEC). FDSs do not belong to vector solitons, as not all the components of FDSs have bright or dark soliton structure. The topological nature of FDSs also clearly distinguish themselves from magnetic solitons.  Moreover, exact solutions for FDSs are available at a strong spin-dependent interaction coupling ($g_s/g_n=-1/2$)~\cite{MDWYuBlair,YuBlairmovingMDW} which is far beyond the Manakov limit ($g_s=0$ in the context of spin-1 BECs).

FDSs are Ising type magnetic domain walls. A natural question is that whether Bloch type and N\'eel type magnetic domain walls could  exist in the easy-plane phase.  A N\'eel type-like domain wall has been investigated numerically in a trapped 1D ferromagnetic BECs at $q=0$~\cite{zhang2007}, which might be related to the wall defects appearing transiently in early quench dynamics from unmagnetized or partially magnetized states~\cite{zhang2005DW,Saito2005DW}. However it is not clear whether the trap potential plays the key role to sustain this structure.  Searching for Bloch type and N\'eel type magnetic domain walls in the easy-plane phase of ferromagnetic BECs deserves future investigations.

So far we only consider excitations from a uniform ground state. In immiscible condensates,   
domain walls are also referred to as the interfaces between the spatially separated components of the condensate, examples are density  (or pseudo magnetization $F_z$) domain walls in immiscible binary BECs~\cite{Ao1998a,Coen2001,chai2020,footnotechai},  and magnetic domain walls in ferromagnetic spin-1 BECs for a negative quadratic Zeeman shift $q<0$~\cite{takeuchi2021easyaxis}.

\section*{Acknowledgment}
X.Y. acknowledges the support from NSAF with grant No. U1930403
and NSFC with grant No. 12175215.  
P.B.B acknowledges support from the Marsden Fund of the Royal Society of New Zealand.


\pagebreak

\end{document}